\documentclass[8pt,aps]{revtex4} 
 
\usepackage{amssymb} 
\usepackage{latexsym} 
\usepackage[dvips]{graphicx} 
\usepackage{epsfig} 
\usepackage{enumerate} 
\usepackage{bigints} 
 
\usepackage{dcolumn} 
\usepackage{amsmath} 
\usepackage{amsfonts} 
\usepackage{bm} 
 
\usepackage{color} 
 
\newcommand{\be}{\begin{equation}} 
\newcommand{\ee}{\end{equation}} 
\newcommand{\bea}{\begin{eqnarray}} 
\newcommand{\eea}{\end{eqnarray}}

\newcommand{\p}{\partial} 
\newcommand{\s}{\sigma} 
\newcommand{\hs}{\hat{\sigma}}

\newcommand{\rd}{\mbox{d}} 
\newcommand{\rcd}{{\cal D}} 
\newcommand{\ri}{\mbox{i}} 
\newcommand{\re}{\mbox{e}}

\renewcommand{\vec}[1]{{\bm #1}}

\begin{document} 
\title{Low Energy Properties of the Kondo chain in the RKKY regime} 
\author{D. H. Schimmel} 
\affiliation{Ludwig Maximilians University, Arnold Sommerfeld Center and Center for Nano-Science, Munich, DE-80333, Germany} 
 
\author{A. M. Tsvelik} 
\affiliation{Department of Condensed Matter Physocs and Material Sciences, Brookhaven National Laboratory, Upton, NY 11973-5000, USA} 
 
\author{O. M. Yevtushenko} 
\affiliation{Institute for Theoretical Physics II, Erlangen, DE-91058, Germany} 
\affiliation{Ludwig Maximilians University, Arnold Sommerfeld Center and Center for Nano-Science, Munich, DE-80333, Germany} 
 
\begin{abstract} 
We study the Kondo chain in the  regime of high spin concentration where the low energy
physics is dominated by the Ruderman-Kittel-Kasuya-Yosida (RKKY) interaction. As has been recently shown (A. M. Tsvelik and O. M. Yevtushenko, Phys. Rev. Lett {\bf 115}, 216402 (2015)), this model has two phases with drastically different transport properties depending on the anisotropy of the exchange interaction. In
particular, the helical symmetry of the fermions
is spontaneously broken when the anisotropy
is of the easy plane type (EP). This leads to a parametrical
suppression of the  localization effects. In the present paper 
we substantially extend the previous theory,
in particular, by analyzing a competition of
forward- and backward- scattering, including
into the theory short range electron interactions
and calculating spin correlation functions. We
discuss applicability of our theory and possible
experiments which could support the theoretical
findings.
\end{abstract} 
 
\date{\today } 
 
\maketitle 
 
\section{Introduction} 
 
The Kondo chain (KC) is one of the archetypal models for interacting low-dimensional
systems which has been intensively studied during the past two decades \cite{review-gulacsi,Loss-Helic1,Loss-Helic2,Gulasch,MaciejkoLattice,KL-Rev1,ZachEmKiv,KL-Honner-Gulacsi,KL-CoulGas,KL-CoulGas-PRL,Shibata}. It consists of  
band electrons on a one-dimensional lattice which interact with localized magnetic moments;  
electron-electron interactions can also be included in the consideration \cite{review-gulacsi,Loss-Helic1,MaciejkoLattice,KL-CoulGas,WhiteAffleck-PRB}. %Such a model  
%is relevant for the theoretical description of certain compounds, e.g. \cite{Ogawa-Cu(pr),Millis_Littlewood_Shraiman,Tokura_Tomioka,lee1986,Suzuki1993347,Zaharko_PRB,PhysRevB.84.180409}.
 The KC is not exactly solvable, nevertheless, a lot is known about it both from numerical 
and analytical studies \cite{ZachEmKiv,KL-Honner-Gulacsi,KL-Rev1,KL-CoulGas,review-gulacsi}.
In particular, ground state properties are known from DMRG for the isotropic point \cite{KL_isotropic_phase_diagram}.

As an example of quasi one-dimensional structures with coexisting  localized and delocalized electrons one may consider the iron-based ladder materials such as $\rm AFe_2 Se_3$ (A= Ba, K and Cs) which crystallize in a structure consisting of ladders formed by edge-shared FeSe$_4$ tetraedra with channels occupied by A atoms. In these materials some of the iron $d$-orbitals are localized and some are itinerant \cite{Dagotto_PRL,PhysRevB.84.180409,Luo2013}. These materials or their modifications may become an experimental realization of the KC model.

It has been recently shown by two of us that the KC may display a rather nontrivial physics in the anisotropic regime away from half-filling in the case of dense spins when the RKKY exchange interaction dominates the Kondo screening  \cite{shortpaper}. We considered   
an anisotropic  exchange interaction with the anisotropy of the XXZ-type. Then there are 
two phases with  different low-energy properties, namely, the Easy Axis phase   
and the Easy Plane one. In the Easy Axis phase, all single fermion excitations  are gapped. The charge transport  
is carried  by collective excitations which can be easily pinned by ever present  
potential disorder. The situation is drastically different in the Easy Plane phase. The minimum 
of the ground state energy corresponds to the helical spin configuration with wave vector $2k_F$ ($k_F$ being the Fermi wave vector)  which  opens a gap in the spectrum of the fermions of a particular  helicity  while the  
electrons having the other (opposite) helicity remain gapless. We remind 
the readers that the helicity is defined as $ {\rm sgn}(v) {\rm sgn} (\sigma)$, where $ v $  
and $ \sigma $ the the electron velocity and its spin, respectively. This corresponds to the 
  spontaneous breaking of the discreet 
$ \mathbb{Z}_2 $ helical symmetry. If the potential disorder is  added to the phase with the broken 
symmetry a single-particle backscattering is prohibited either by spin conservation 
(for  electrons with the same helicity) or by the gap in one of the helical sectors 
(for  electrons with  different helicity). This is similar to the absence of the single-particle 
back-scattering of edge modes in time-reversal invariant topological insulators \cite{MaciejkoLattice,MooreBalents,MaciejkoOregZhang,RoyZ2,XuMoore,MolFranz,Kurita_Yamaji_Imada_TI,Kawakami_Hu_HelicalEdge}
and results in suppression of localization effects. The latter can appear only due to collective  
effects resulting in a parametrically large localization radius. In other words, ballistic charge 
transport in the EP phase has a partial symmetry protection which is removed either 
in very long samples or if the spin U(1) symmetry is broken. This is also similar to the symmetry protection of the edge transport in 2d topological insulators: transport is ideal if time-reversal symmetry and spin $U(1)$ symmetry are present. However, it can be suppressed in a long sample due to spontaneously broken time-reversal symmetry \cite{AAY,Yevt-Helical}. 
 
In the present paper, we continue to study  the KC in the RKKY regime where the 
low energy physics is governed by the fermionic gaps. We aim to explain in more details  
the results of Ref.\cite{shortpaper} and to substantially extend the theory, in particular, by analyzing  
the role of forward scattering (i.e., of the Kondo physics), by taking into account the short  
range electron interactions and by calculating the spin correlation functions. 

Similar ideas to those presented here were already pursued in \cite{Loss-Helic1}, where the emergence of helical order was recognised. 
In contrast to \cite{Loss-Helic1} we take into account the dynamics of the lattice spins whose presence substantially modifies the low-energy theory. 
 
The Hamiltonian of the KC on a lattice is 
\be 
	{\cal H} = {\cal H}_0 + {\cal H}_{\rm int} = \sum_{i} \Big[t c^\dagger_{i+1}  c_i + H.c. \Big] + \sum_a \sum_{j \in M} J^a S^a_j c^\dagger_j \sigma^a c_j, \ 
                a = x,y,z ;
\label{model} 
\ee 
where $t$ is the hopping matrix element, $c_i^{(\dagger)}$ annihilates (creates) an electron at site $i$, $\vec S_i$ is a local spin of magnitude $s$,  
$\sigma^a$ is a Pauli matrix, and $M$ constitutes a subset of all lattice sites. $J$ denotes the interaction strength between the impurities  
and the electrons. We distinguish $J_z$ and $J_x = J_y =: J_\perp$. Short range interactions between the electrons will be added later in section \ref{sec:interactions}.
The dynamics of a chain of spins will be added in section \ref{sec:WZ}.
We will be interested in the case of dense magnetic impurities, $\rho_s \gg 1/L_K$ (with the impurity density $\rho_s$ and the single-impurity Kondo length $L_K$), when the effects of the electron-induced exchange can 
take predominance over the Kondo screening.  
 
The paper is organized as follows: 
We first introduce a convenient representation of the impurity spins in section \ref{sec:WZ}.
Necessary conditions for the RKKY regime are then discussed in Section \ref{sec:scattering_RG}.  
The gap is studied in section \ref{sec:BS}.
In section \ref{sec:Density_correlations} we compute the conductance and analyze the effects of spinless disorder.
The spin-spin correlation functions are given in section \ref{sec:spins}.

\section{Formulation of the low energy theory} 
\label{sec:WZ}
\begin{figure} 
\includegraphics{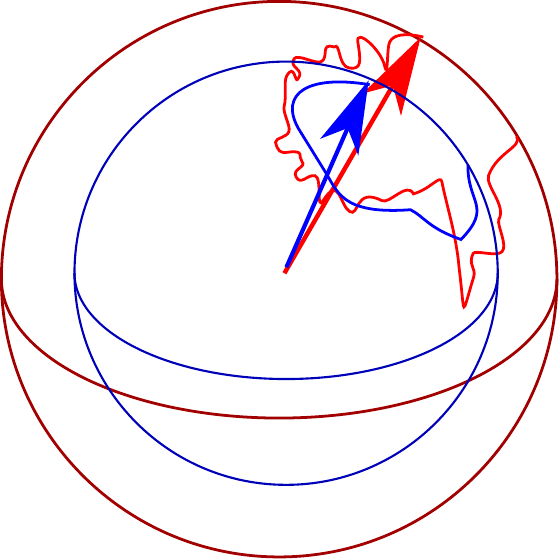} 
\quad \quad \quad \quad \includegraphics{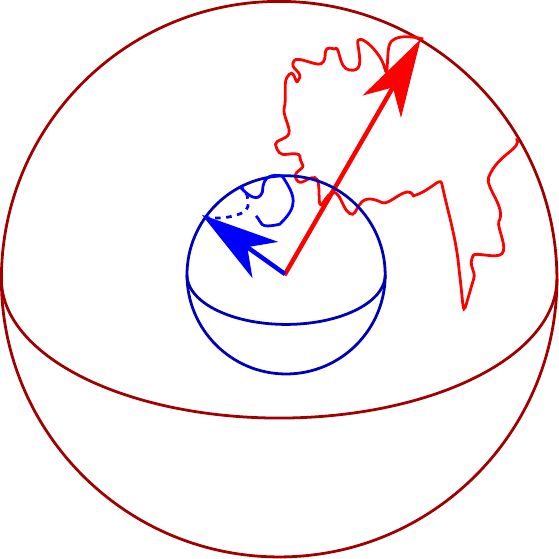} 
\caption{(color online) The fast (red) and slow (blue) spin trajectories as a function of time. The slow spin is shorter, since it is the fast spin averaged over some short timescale. The left panel shows the slow modes in the case of a free spin, in the right panel the spin physics is dominated by the interaction mediated by the backscattering electrons. In the latter case, the slow mode is orthogonal to the fast trajectory.} 
\label{fig:trajectories} 
\end{figure} 
\begin{figure} 
\includegraphics[scale=.3]{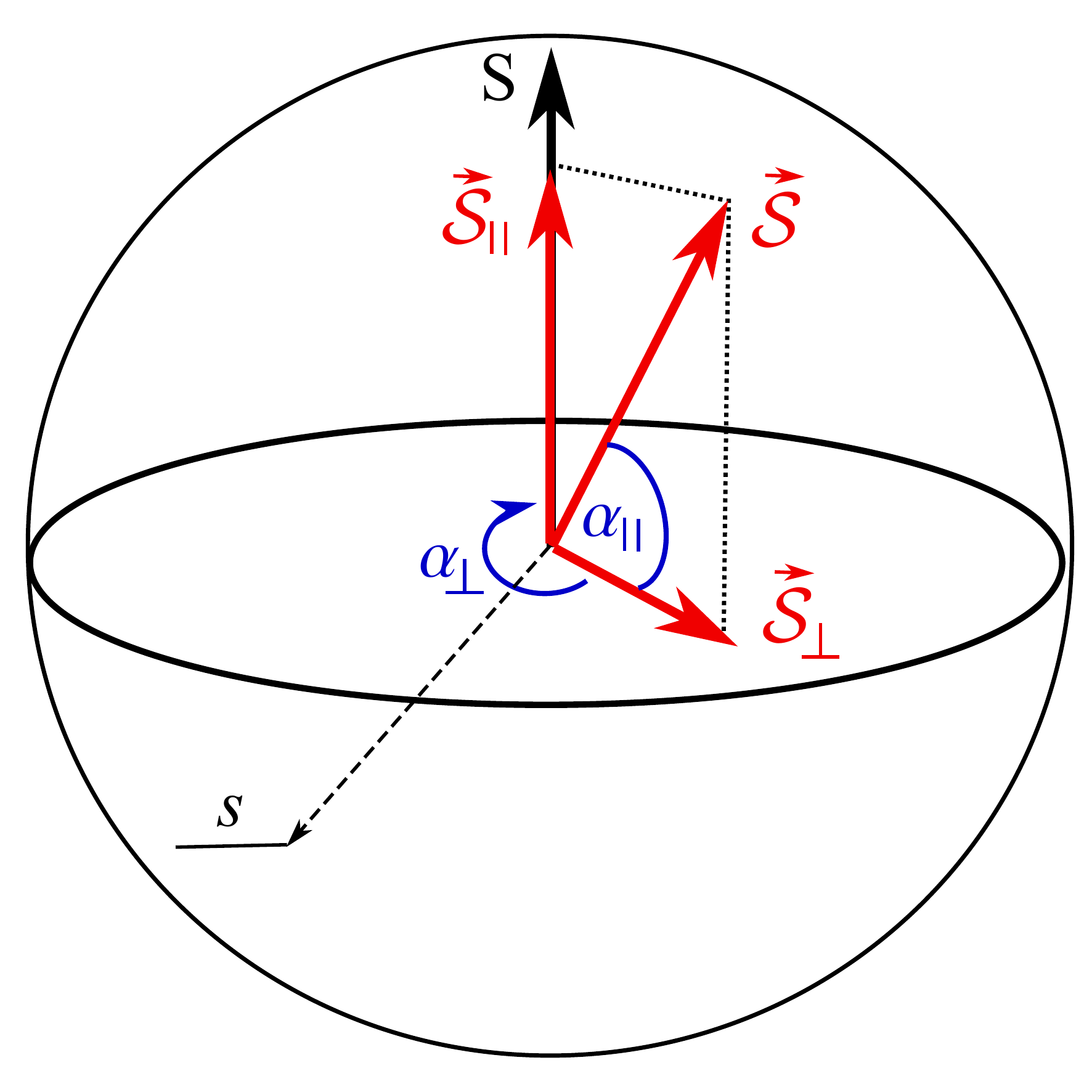} 
\caption{(color online) The parametrization of a spin by the angles $\theta$, $\psi$, $\alpha_\perp$, $\alpha_\parallel$.} 
\label{fig:geometry} 
\end{figure} 
 
To develop a low energy description of the KC model (\ref{model}) we have to single out slow modes and integrate over the fast ones. 
As the first step, we need to find a convenient representation of the spins such that it will be easy to separate the low and high energy degrees of freedom. 
 
\subsection{Separation of scales in the spin sector} 
Consider first a single spin. It is described by the Wess-Zumino term in the action \cite{ATsBook} 
\be 
	S_{WZ} = \ri\int_0^1 d u \int_0^\beta d\tau \frac{s}{8\pi} \epsilon^{\mu \nu} \vec n \cdot (\partial_\mu \vec n \times \partial_\nu \vec n), 
\ee 
where $\vec n$ is the direction of the spin, $u$ is an auxiliary coordinate, which together with $\tau$ parametrizes a disk.  
Multiple spins require a summation over spins and can be described by introducing a (dimensionless) spin density $\rho_s$ 
\be  \label{WZ-basic1} 
   \sum_{\rm impurities} S_{WZ} \rightarrow S = \int dx \frac{\rho_s}{\xi_0} S_{WZ}, 
\ee 
where $\xi_0$ %can be associated with an  
is the underlying lattice constant for the spins. 
 
Usually, two angular variables are used in parametrizing the spin  
$ {\bf S} = s \{ \sin(\theta)\cos(\psi), \sin(\theta)\sin(\psi), \cos(\theta) \} $: 
\be  \label{WZ-basic} 
   {\cal L}_{\rm WZ}[\theta,\psi] =  \frac{\ri s \rho_s}{\xi_0} \cos\theta\p_{\tau} \psi \,  , 
\ee 
where we have neglected boundary contributions (topological terms).  
 
The form of the Lagrangian Eq. \eqref{WZ-basic} makes it difficult to separate fast and slow variables, since the angles $\theta$ and $\psi$ contain both fast and slow modes. We need to find a different representation of the spin Berry phase, which will allow us to separate the fast and the slow modes explicitly.  
We first observe that the expression Eq. \eqref{WZ-basic} can be obtained by considering a coordinate system comoving with the spin. Namely, we choose an orthnormal basis $\{ \vec e_1, \vec e_2, \vec e_3 \}$ at time $\tau = 0$ and assume that this coordinate system is comoving with the spin such that $ s^e_i := (\vec S, \vec e_i) $ is independent of $\tau$.  
Then it is easy to check that the following expression reproduces (\ref{WZ-basic}): 
\be 
	\label{WZ-inv} 
	{\cal L}_{\rm WZ}[\theta,\psi] = - \frac{\ri \rho_s}{2 \xi_0} ({\bf S}, {\bf e}_i) (  {\bf e}_j,  \partial_\tau {\bf e}_k ) \epsilon_{ijk}. 
\ee 
 
The check of Eq. (\ref{WZ-inv}) can be done by choosing the explicit parametrization %considering the basis 
\begin{subequations} \label{WZ-basis1} 
\bea 
   {\bf e}_1 &=& \{ -\cos(\theta) \cos(\psi), -\cos(\theta) \sin(\psi), \sin(\theta) \} , \\ 
   {\bf e}_2 &=& \{ \sin(\psi), -\cos(\psi), 0\} , \\ 
	{\bf e}_3 &=& \{ \sin(\theta)\cos(\psi), \sin(\theta)\sin(\psi), \cos(\theta) \} = \vec S/s, 
\eea 
\end{subequations} 
with $\vec S \parallel {\bf e}_3$ and inserting Eq. \eqref{WZ-basis1} into Eq. \eqref{WZ-inv}. 
A specific choice of the basis ${\bf e}_{2,3}$ is not important since ${\cal L}_{\rm WZ}$ in the form Eq. \eqref{WZ-inv} is manifestly covariant under both a rotation in $x$, $y$, $z$, and a change of basis $\{{\bf e}_i\}$.  
 
In path integral quantization, we thus sum over all paths described by $\theta(x, \tau)$ and $\psi(x, \tau)$. The measure is given by $\rcd \{\Omega\} = \sin \theta \rcd \{ \theta \} \rcd \{ \psi \} $.
 
Let us now consider two superimposed spin motions: the actual trajectory considered in the path integral, and its slow component (Fig. \ref{fig:trajectories}). We already have the Wess-Zumino term for the actual trajectory. If we want to use Eq. \eqref{WZ-inv} for the slow component, we need to introduce a second set of basis vectors which is comoving with the slow component. This doubles the number of angles, but we assume a separation of scales: of the four angles, two will be fast and two will be slow. Thus, there will be no double counting of modes which justifies our approach. 
A convenient choice for the slow basis is given by the rotation of the actual trajectory (Fig. \ref{fig:geometry})
\begin{subequations} \label{WZ-basis2} 
\bea 
 {\vec e}_1' & = & -\sin({\alpha_\parallel}) [\cos( \alpha_\perp ) {\bf e}_1 + \sin( \alpha_\perp ){\bf e}_2] +  \cos({\alpha_\parallel}) {\bf e}_3, \\ 
 {\vec e}_2' & = & \sin( \alpha_\perp ) {\bf e}_1 - \cos( \alpha_\perp ){\bf e}_2,\\ 
 {\vec e}_3' & = & \cos({\alpha_\parallel}) [\cos(\alpha_\perp) {\bf e}_1 + \sin(\alpha_\perp){\bf e}_2]  + \sin({\alpha_\parallel}) {\bf e}_3 . 
\eea 
\end{subequations}  
The total path-integral measure now consists of the four angles: $\rcd \{\Omega_S, \Omega_{S'}\} = \cos \alpha_\parallel \sin \theta \rcd \{\theta\} \rcd \{\psi\} \rcd \{\alpha_\parallel\} \rcd \{\alpha_\perp\} $, which will be the product of the measures for fast and slow modes. 
 
Now we can describe the dynamics of the slow modes, which is given by the slow Wess-Zumino term: we pick the bases such that ${\bf S} \parallel {\bf e_3}$ and ${\bf S_{slow}} \parallel {\bf e_3'}$. The dynamics of the slow modes are then obtained by using Eq. \eqref{WZ-inv} with the full spin ${\bf S}$ and the slow basis ${\bf e_3'}$: 
\bea 
   \label{WZ} 
  S_{\rm WZ}^{slow} = %\int \rd x  {\cal L}^{slow}_{\rm WZ} %& \approx & - \frac{\ri \rho_s}{2 \xi_0} ({\bf S}, {\bf e}_i') (  {\bf e}_j',  \partial_\tau {\bf e}_k' ) \epsilon_{ijk} \\ 
                            & = & \ri s \rho_s \xi_0^{-1} \int \rd x \int \rd t \sin(\alpha_\parallel) [  \p_\tau \alpha_\perp + \cos(\theta) \p_\tau\psi  ] .
\eea 
The dynamics is that of the basis $\{ \vec e_1', \vec e_2', \vec e_3' \}$ (i.e. of the slow spin), whereas the overall scale is that of the actual trajectory projected onto the slow component. This projection may be viewed as a renormalization of the length of the spin's slow component.

\subsection{The interaction between the spins and the fermions} 
The low-energy fermion modes are obtained by linearizing the spectrum and expanding the operators $\hat c$ in smooth chiral modes $\hat R_\sigma$, $\hat L_\sigma$ 
\be 
  \label{fermions} 
  \hat{c}_{ \uparrow \downarrow}(n) = \re^{-\ri k_F \xi_0n} \hat{R}_{ \uparrow \downarrow}(x) + \re^{\ri k_F \xi_0n}\hat{L}_{ \uparrow \downarrow}(x), \, \,  x = n \xi_0. 
\ee 
The Lagrangian density of the band electrons becomes 
\be 
\label{BandLagr} 
  {\cal L}_{\rm e} = \Psi^\dagger \Big[ (\hat{I} \otimes \hat{I} )\p_{\tau} - \ri ( \hat{I} \otimes \hat{\tau}^z) v_F \p_x \Big]\Psi . 
\ee 
The first space in the tensor product is the spin one, the Pauli matrices $ \hat{\tau}^a$ act in the chiral space; $ \hat{I} = {\rm diag}(1,1) $; $ \, v_F = 2 t  \xi_0 \sin( k_F \xi_0 ) $ is the Fermi velocity; $ \, \Psi^{\rm T} = \left( R_\uparrow, R_\downarrow, L_\uparrow, L_\downarrow \right) \, $ is the 4-component fermionic spinor field.  
If the electron interaction is taken into account, it is more convenient to use the bosonized Lagrangian density  
\be 
\label{BandLagrBos} 
  {\cal L}_{\rm e} = - \sum_{\rho = \rm c,s} \left \{\frac{\ri}{\pi} \partial_x \Theta_\rho \partial_\tau \Phi_\rho - \frac{1}{2\pi} \Big[u_\rho K_\rho (\partial_x \Theta_\rho)^2 + \frac{u_\rho}{K_\rho} (\partial_x \Phi_\rho)^2\Big]\right \} , 
\ee 
where $K_\rho$ is the Luttinger paramter; $u_\rho$ the renormalized Fermi velocity; and we have used the bosonization identity 
\be 
	\psi_{r \sigma} = \frac{1}{\sqrt{2\pi \xi_0}} U_{\sigma} \re^{-\ri r k_F x} \re^{-\frac{\ri}{\sqrt 2} [r \Phi_c-\Theta_c+\sigma(r \Phi_s-\Theta_s)]}. 
\ee 
$\Phi_c$ ($\Phi_s$) and $\Theta_c$ ($\Theta_s$) are dual bosonic fields belonging to the charge (spin) sector, $r$ distinguishes right- and left-moving modes, $\sigma$ is the spin projection and $U_{\s}$ are Klein factors. 
One can introduce spin and charge sources to determine how the low energy degrees of freedom couple to external perturbations: 
\be 
\label{Sources} 
  {\cal L}_{\rm source} = h_c (\rho_c^R + \rho_c^L) + h_s (\rho_s^R-\rho_s^L) = -\frac{\sqrt 2 h_c}{\pi} \partial_x \Phi_c + \frac{\sqrt 2 h_s}{\pi} \partial_x \Theta_s, 
\ee 
%here $\rho_{c(s)}^{R/L}$ is the charge (spin) density of the right-/left-moving electrons. 
here $\rho_{c/s}^{R/L} = \rho_\uparrow^{R/L}\pm \rho_\downarrow^{R/L}$ is the charge/spin density of the right-/left-moving electrons.
%and $\rho_{s}^{R/L} = \rho_\uparrow^{R/L}-\rho_\downarrow^{R/L}$ is the spin density of the right-/left-moving electrons.
The spin source is included for purely illustrative purposes. 
We will combine the fermionic and bosonic description, selecting the one which is most convenient for the given caculations. 
 
Now consider the electron-spin interactions ${\cal H}_{int}$.
%\be 
%   \hat{H}_{int} = \sum_m J_a \, \hat{c}^\dagger_m \, \hs^a \hat{S}^a(m) \, \hat{c}_m . 
%\ee 
We will explicitely distinguish forward and backward scattering since they give rise to different physics. 
The slow part of the backscattering term is (c.f. Appendix) 
\be 
\label{CouplLagrSlowParam} 
   {\cal L}_{int}^{\rm (sl, bs)} = \frac{s \cos(\alpha_\parallel) \rho_s}{2} 
             R^\dagger \Bigl\{ 
                 J_\perp \left[ \re^{\ri \psi} \sin^2 \! \left( \frac{\theta}{2} \right)\hs^-  \!\! - 
                                       \re^{-\ri \psi} \cos^2 \! \left( \frac{\theta}{2} \right)\hs^+ 
                               \right] 
                                  + 2 J_z \sin(\theta)\hs^z \Bigr\} L \re^{-\ri \alpha} + H.c., 
\ee 
where $\alpha = \alpha_\perp - 2 k_F x$ and we have introduced the spin-flip operator $S_\pm = S_x \pm iS_y$. 
 
For the forward scattering, we obtain 
\be 
\label{CouplLagrSlowParamForward} 
	{\cal L}_{int}^{\rm (sl, fs)} = \frac{s \sin(\alpha_\parallel) \rho_s}{2} R^\dagger \left \{ J_\perp^f \sin \theta \left [\re^{\ri \psi} \sigma^- + \re^{-\ri \psi} \sigma^+ \right ] + 2 J_z^f \cos \theta \sigma^z \right \} R + (R \rightarrow L) 
\ee

\section{Renormalization of forward vs backward scattering coupling constants} 
\label{sec:scattering_RG} 
 
Eq. \eqref{CouplLagrSlowParam} and Eq. \eqref{CouplLagrSlowParamForward} describe two competing phenomena: forward scattering tends towards Kondo-type physics, backward scattering opens a gap (c.f. section \ref{sec:BS}). Both phenomena are distinct and mutually exclusive. If backscattering is dominant, then the emerging gap will cut the RG and suppress forward scattering. If forward scattering dominates, the formation of Kondo-singlett prevents the gap from opening \cite{ZachEmKiv}. We will focus on the physics related to the gaps. Therefore, we have to identify conditions under which the backscattering terms are more important. To determine the dominant term, we consider a first loop RG.  
 
Let us consider the bosonized free electrons, Eq. \eqref{BandLagrBos}. They constitute two Luttinger liquids, describing a spin density wave (SDW) and a charge density wave (CDW). If there is no electron-electron interaction, then $K_s = K_c = 1$. A weak, short range, spin independent repulsion between electrons changes $K_c$ to $K_c \lesssim 1$, but leaves $K_s$ untouched. 
 
The RG equations for the couplings read as (see Appendix \ref{sec:App_RG}): 
\bea 
	\partial_l J_z^f = - J_z^f,                                          & \partial_l J_\perp^f = \left [ \tfrac 12 (K_s + \frac{1}{K_s}) - 2 \right ] J_\perp^f, \\ 
	\partial_l J_z^b = \left [ \tfrac 12 (K_s + K_c) - 2 \right ] J_z^b, & \partial_l J_\perp^b = \left [ \tfrac 12 (K_c + \frac{1}{K_s}) - 2 \right ] J_\perp^b, 
\eea 
where $l$ parametrizes an energy cutoff $\Lambda'$ via $\Lambda' = \exp(l) \Lambda$. The flow differs from that of single Kondo impurity because we consider a \textit{dense array} of impurities.  
All of these terms are relevant, if $K_c$ and $K_s$ are close to $1$. Assuming weak, short range, spin independent repulsion (i.e. $K_c \lesssim 1$, and $K_s = 1$), we see that the backward scattering terms flow faster in the RG-flow from high to low energies than forward scattering ones, i.e. the terms $\sim J^b$ can dominate. 
 
Let us assume that an impurity scatters anisotropically in spin space ($J_z \neq J_\perp$), but there is no difference between the electrons' directions ($J^f_{\rm bare} = J^b_{\rm bare}$). Then, simple scaling shows that backward scattering becomes relevant prior to forward scattering. The scattering will remain anisotropic and the strength of the anisotropy is dictated by the inital conditions ($J_z$ vs. $J_\perp$ at the beginning of the flow). 
 
Weak, short range, spin dependent electron-electron interactions do not change the picture and backscattering dominates, provided that $\vert K_s -1 \vert < \vert K_c -1 \vert$. However, if the spin dependent electron-electron interactions are attractive (repulsive), they will drive the flow towards dominantly spin-flip (spin-conserving) backscattering. 
 
Thus, we conclude that the gap physics dominates if there is a weak, repulsive, spin-independent electron-electron interaction. From now on, we consider this regime and neglect $J^f$. 
We note that it is well-known that for large spins the Kondo-temperature is small \cite{Schrieffer_large_spin}. Thus, for sufficiently large spins we can conclude without an explicit RG analysis that the gap physics will dominate.

\section{Effects of backward scattering} 
\label{sec:BS} 
\begin{figure} 
\includegraphics[scale=.5]{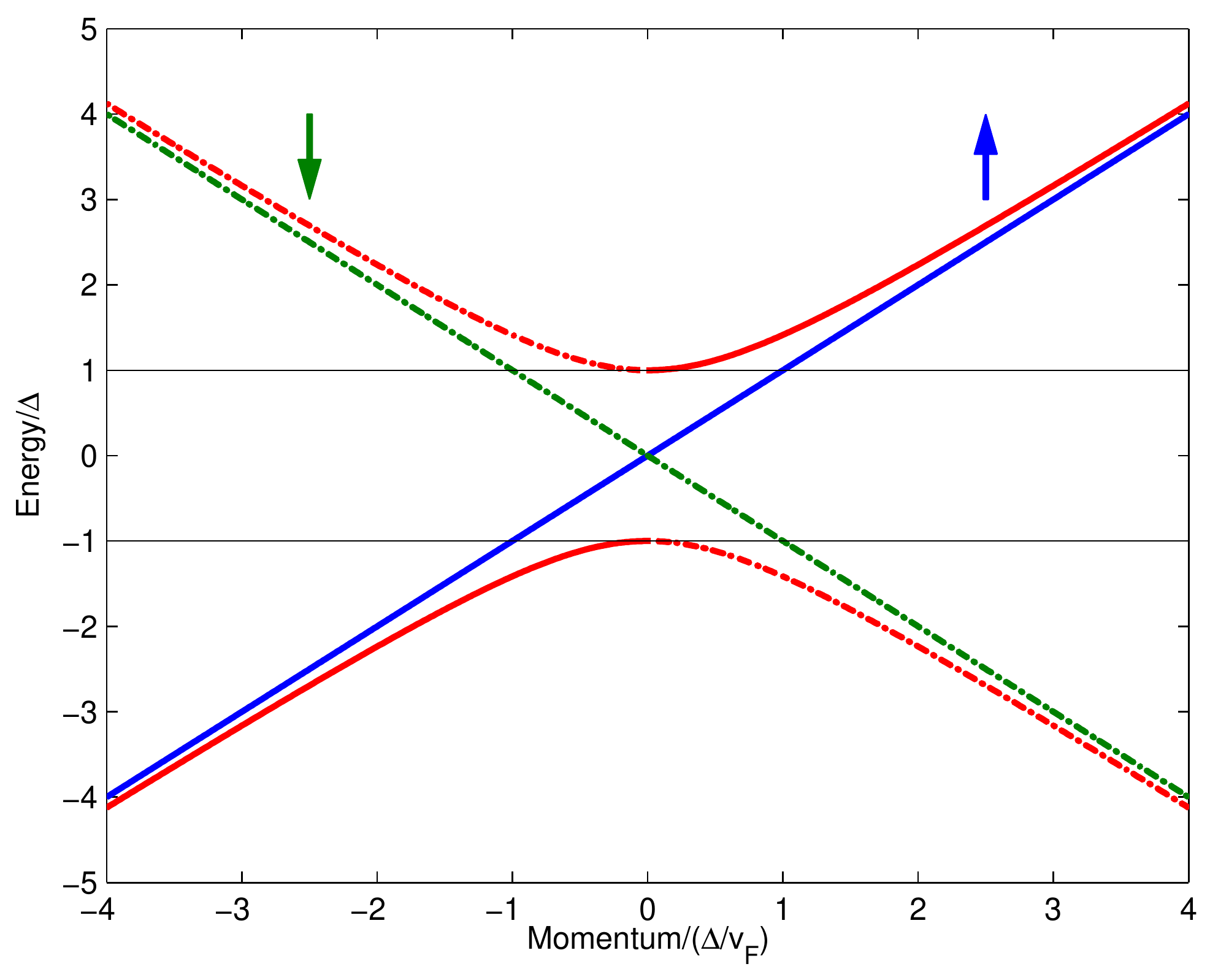} 
\caption{(color online) The dispersion of helical modes. Blue and green lines correspond to particles and holes of the first helical sector. For helical particles the direction is in one-to-one correspondes with their spin. Upon opening a gap $\Delta$, the dispersion changes to the red curve.} 
\label{fig:gap} 
\end{figure}

We now focus on effects generated by backscattering. If the spin configuration is fixed, the backscattering terms act like mass terms for the fermions. This modifies the dispersion relations, as shown in Fig. \ref{fig:gap}.
The ground state energy of single component massive fermions with mass $m$ differs from that of gapless fermions by 
\be 
	\label{GSEnergy-Dirac} 
	\Delta E = - \frac{\xi_0}{2 \pi v_F} m^2 \ln (t/\vert m \vert) + {\cal O} (m^2). 
\ee 
%{\color{red} Derive GS energy. } 
To minimize the ground state energy, one thus has to maximize the gaps. Depending on the relative values of $J^z$ and $J^\perp$ this leads to different ground state spin configurations and different physics.

\subsection{Easy axis anisotropy, $J_z \gg J_\perp$} 

\label{sec:easy_axis} 
Let us consider  $J_z \gg J_\perp$. It is convenient to remove the phases $\alpha$ and $\psi$ from  the interaction Eq. \eqref{CouplLagrSlowParam}. This can be done by the transformation of the fermion fields  
\be 
\label{fermion-trafo} 
	R_\uparrow \rightarrow \re^{-\ri \psi/2-\ri \alpha/2} R_\uparrow, \quad R_\downarrow \rightarrow \re^{\ri \psi/2-\ri \alpha/2} R_\downarrow, \quad L_\uparrow \rightarrow \re^{-\ri \psi/2+\ri \alpha/2} L_\uparrow, \quad L_\downarrow \rightarrow \re^{\ri \psi/2+\ri \alpha/2} L_\downarrow, 
\ee 
which is anomalous. The anomaly is the well-known Tomonaga-Luttinger anomaly; its contribution to  the Lagrangian is \cite{Yurkevich-2004} 
\be 
\label{action_TL}
	\sum_{\sqrt 2 \Phi = \alpha, \psi} {\cal L}_{TL}[\Phi, v_F] = \sum_{\sqrt 2 \Phi = \alpha, \psi} \frac{1}{2 \pi v_F} \left[ (\partial_\tau \Phi)^2 + (v_F \partial_x \Phi)^2 \right]. 
\ee 
This result may also be obtained from Abelian bosonization \cite{Bosonisation} (see the Appendix \ref{sec:App_Shift}) \footnote{We use the conventions from Ref. \cite{Giamarchi}}. 
We have neglected coupling between the charge (spin) density and the field $\alpha$ ($\psi$). 
This mixing is generically of the form 
\be 
\label{L_mixing} 
	{\cal L}_{\rm mixing} \sim \ri \partial_\tau \alpha (\rho_L-\rho_R)_c + u \partial_x \alpha (\rho_L+\rho_R)_c + \ri \partial_\tau \psi (\rho_L + \rho_R)_s + u \partial_x \psi (\rho_L - \rho_R)_s, 
\ee 
where $\rho_r$ stands for a density of left-/right-moving ($r=L$ and $r=R$) electrons and $u$ is their velocity. Once the electrons become gapped, the low-energy degrees of freedom  cannot excite density fluctuations. With this accuracy, in the low energy theory we can neglect derivatives of the electron densities. 
 
The full Lagrangian is thus 
\be 
	\label{Ltotal} 
  {\cal L}^{\rm (sl)} \simeq {\cal L}_{\rm e} + {\cal L}_{int}^{\rm (sl)} \! |_{\alpha,\psi=0} + \sum_{\sqrt 2 \Phi=\alpha,\psi} \!\!\!\! {\cal L}_{\rm TL}(\Phi,v_F) + {\cal L}_{\rm WZ}; 
\ee 
Here ${\cal L}_{int}^{\rm (sl)}$ is only the backward scattering part ${\cal L}_{int}^{\rm (sl,bs)}$, Eq. \eqref{CouplLagrSlowParam}.
After the transformation Eq. \eqref{fermion-trafo}, the sources now couple to the phases $\Phi_c$ and $\Theta_s$ and the angles 
\be 
\label{Sources2} 
  {\cal L}_{\rm source} = -\frac{h_c}{\pi} \partial_x \alpha - \frac{h_s}{\pi} \partial_x \psi -\frac{\sqrt 2 h_c}{\pi} \partial_x \Phi_c + \frac{\sqrt 2 h_s}{\pi} \partial_x \Theta_s. 
\ee 
${\cal L}_{int}^{\rm (sl)}$ in Eq. \eqref{Ltotal} is a mass term. The masses for fixed spin variables are given by 
\be 
\label{GapsPM} 
   m_{\pm}^2 = \frac{ (s \cos \alpha_\parallel \rho_s )^2}{4} \left( \sqrt{(J^b_\perp)^2 \cos^2\theta + (J^b_{z})^2\sin^2\theta} \pm J^b_\perp \right)^2. 
\ee 
In the case of $J_z \gg J_\perp$ the gap is always large (of order $J_z$) and it is maximized for $\theta=\pi/2$ and $\alpha_\parallel=0$.  
 
Since all fermions are gapped we may neglect their coupling to external sources, provided we restrict ourselves to energies below the gap. We now integrate out the fermions under this assumption, i.e. we will consider correlation functions on length scales larger than the coherence length $v_F/m$. Since the original normalization of the path integral was with respect to gapless fermions, the effective Lagrangian is now changed by the fermionic ground state energy Eq. \eqref{GSEnergy-Dirac}.  
The total Lagrangian reads as 
\be 
	\label{eq:L_ea_no_electrons} 
  {\cal L}^{\rm (sl)} \simeq -\frac{\Delta E}{\xi_0} + \sum_{\sqrt 2 \Phi=\alpha,\psi} \!\!\!\! {\cal L}_{\rm TL}(\Phi,v_F) + {\cal L}_{\rm WZ}, 
\ee 
where we also have assumed that fluctations of the angles $\theta$ and $\alpha_\parallel$ are small, such that the angles are close to their ground state values. $\Delta E$ is a function of the angles, see equations \eqref{GSEnergy-Dirac} and \eqref{GapsPM}. Expanding Eq. \eqref{eq:L_ea_no_electrons} in $\theta' = \theta - \pi/2$ and $\alpha_\parallel$, we obtain 
 
\be 
\label{L_sl_ea_1} 
	{\cal L}^{(sl)}_{(ea)} = \sum_{\sqrt 2 \Phi=\alpha,\psi} \!\!\!\! {\cal L}_{\rm TL}(\Phi,v_F) + \underbrace{a \left \{ \left [ ({J_z^b})^2 - ({J_\perp^b})^2 \right ] (\theta')^2 + \left [ ({J_z^b})^2 + ({J_\perp^b})^2 \right ] (\alpha_\parallel)^2 \right \} }_{{\cal L}_{gs}} + \ri s \rho_s \xi_0^{-1} \alpha_\parallel (\partial_\tau \alpha), 
\ee 
where $a = \log(t/J) (s\rho_s)^2/4\pi v_F$, and we do not distinguish between the $J$'s in the $\log$. We will further assume for now that $\partial_\tau \psi$ is small, such that the cross-term $\alpha_\parallel \theta' \partial_\tau \psi$ is a higher order contribution. This will be verified below. ${\cal L}_{\rm gs}$ in Eq. \eqref{L_sl_ea_1} is the mass term for $\theta'$ and $\alpha_\parallel$, which shows that the assumption of small $\theta'$ and $\alpha_\parallel$ is consistent. 
 
Now we perform the integrals over $\alpha_\parallel$ and $\theta'$ and obtain 
\be 
\label{L_ea} 
	{\cal L}^{(sl)}_{(ea)} = \sum_{\sqrt 2 \Phi=\alpha,\psi} \!\!\!\! {\cal L}_{\rm TL}(\Phi,v_F) + \frac{(s \rho_s \xi_0^{-1})^2}{4 a (({J_z^b})^2 + ({J_\perp^b})^2)} (\partial_\tau \alpha)^2 
\ee 
Note that $\psi$ and $\alpha$ remain gapless, justifying the previous approximation of small $\partial_\tau \psi$.  
Thus, two angular modes are fast ($\theta$ and $\alpha_\parallel$) and two are slow ($\alpha$ and $\psi$), as we expected. 
 
Eq. \eqref{L_ea} is the action of two $U(1)$-symmetric Luttinger liquids with a charge mode, $\alpha$, and a spin mode, $\psi$. 
\be 
\label{LagrEA} 
   {\cal L}_{\rm ea} = \frac{ 1 }{ 2 }{\cal L}_{\rm TL}(\psi,v_F) + \frac{ 1 }{ K_\alpha } {\cal L}_{\rm TL}( \alpha,v_\alpha ) \, . 
\ee 
The two phases couple to different sources: $\alpha$ to charges and $\psi$ to spins. 
The slow mode $\alpha$ has a renormalized velocity and Luttinger parameter 
\be 
\label{Compr-a} 
  \frac{v_\alpha}{v_F} = \frac{K_\alpha}{2} \simeq \xi_0 \frac{\sqrt{J_z^2 + J_\perp^2}}{\pi v_F } \sqrt{\log(t/J)} \ll 1 , 
\ee 
where we used that the band width is the largest energy scale (i.e. $v_F /\xi_0 \gg J$) in the last inequality. This severly affects the charge transport, which is mediated by $\alpha$.

\subsection{Breaking the $\mathbb{Z}_2$ symmetry} 
We have demonstrated that for $J_z \gg J_\perp$, all fermionic modes have approximately the same gap $\sim J_z$. Approaching the $SU(2)$ symmtric point, the mass $m_-$ shrinks until it would reach zero at $J_z = J_\perp$. In terms of the easy axis (EA) picture, some fermions (two helical modes) become light and their contribution encompasses large fluctuations on top of their ground state energy. We explicitely assumed that the fluctuations around the ground state are small. Therefore, our approach is no longer valid for $m_- \rightarrow 0$.  
 
For now, let us consider the other limit $J_z \ll J_\perp$. 
We will see that this parameter regime behaves in a way qualitatively different to $J_z \gg J_\perp$.  
%We will show the existence of a quantum phase transition, which separates a phase with broken $\mathbb Z_2$ symmtry between the helical sectors from a phase where this symmtry is unbroken.  
The order parameter distinguishing the phases is discussed in section \ref{sec:spins}. 
The vanishing of the gap for $J_z \rightarrow J_\perp$, the spontaneous symmetry breaking for $J_z \ll J_\perp$ and the presence of an order parameter all strongly suggest the presence of a quantum phase transition, although its theoretical description is missing.

\subsection{Easy plane anisotropy, $J_z \ll J_\perp$} 
\label{sec:easy_plane} 
Let us put for simplicity $J_z \rightarrow 0$. Then, it is convenient to express Eq. \eqref{CouplLagrSlowParam} through helical modes 
\bea 
\label{Helic-1} 
   {\cal L}_{\rm bs}^{\rm (h_1)} & = & s \cos \alpha_\parallel \rho_s\Big[ 
            J_\perp R_{\uparrow}^\dagger \cos^2\left( \theta/2 \right) \re^{-\ri (\psi+\alpha)} L_{\downarrow} + H.c.\Big], \\ 
\label{Helic-2} 
   {\cal L}_{\rm bs}^{\rm (h_2)} & = & - s \cos \alpha_\parallel \rho_s\Big[ 
            J_\perp R_{\downarrow}^\dagger \sin^2\left( \theta/2 \right) \re^{\ri (\psi-\alpha)} L_{\uparrow} + H.c.\Big] .
\eea 
Clearly, the interesting points are $\theta = 0, \pi$ and $\theta = \pi/2$. If $\theta=\pi/2$, then the effective $J_\perp$ is reduced by a factor of $\cos^2 \pi/4 = \sin^2 \pi/4 = \tfrac 12$ relative to the effective $J_\perp$ of a single gapped helical sector at $\theta = 0, \pi$. 
Since the ground state energy Eq. \eqref{GSEnergy-Dirac} of a helical sectors with the gap $m_i$ is 
\be 
	\Delta E_{hel} = - \frac{\xi_0}{2 \pi v_F} m_i^2 \ln (t/\vert m_i \vert) + {\cal O} (m_i^2), \quad m_i \sim J_\perp, 
\ee 
the ground state of a single gapped sector of twice the mass has a lower energy than that of two equally gapped helical sectors. 
Thus, it is energetically favorable to spontaneously break the $\mathbb Z_2$ symmetry between different helical sectors. The two ground states are labelled by $\theta = 0$ and $\theta=\pi$. 
 
Let us choose $\theta =0$. Then, the first helical sector Eq. \eqref{Helic-1} becomes gapped, while the second sector Eq. \eqref{Helic-2} is gapless. Now, the angle $\alpha-\psi$ does not enter the action if fluctuations of $\theta$ are set to zero. It enters (in the leading order in $\theta$) only via the combination 
\be 
	{\cal L} \supset \underbrace{-s \cos \alpha_\parallel \rho_s J_\perp \frac{\theta^2}{4} \re^{-\ri (\alpha -\psi)} R^\dagger_\downarrow L_\uparrow + H.c.}_{{\cal L}_{\rm bs}^{\rm (H2)}} + \underbrace{\ri s \rho_s \xi_0^{-1} \sin (\alpha_\parallel) \frac{\theta^2}{2} \partial_\tau \alpha}_{{\cal L}_{\rm WZ}^{slow}}. 
\ee 
The last summand is (for $\alpha_\parallel \approx 0$) beyond our accuracy and will be neglected. The influence of the first two summands may be estimated by integrating over $R_\downarrow$ and $L_\uparrow$. The resulting expression is  
\be 
	{\cal L} \supset Tr \log \left[ \begin{pmatrix} \ri \omega + v_F k & 0 \\ 0 & \ri \omega - v_F k\end{pmatrix}^{-1} 
	                         \begin{pmatrix} \ri \omega + v_F k & s \cos \alpha_\parallel \rho_s J_\perp \frac{\theta^2}{4} \re^{-\ri (\alpha-\psi)} \\ s \cos \alpha_\parallel \rho_s J_\perp \frac{\theta^2}{4} \re^{\ri (\alpha-\psi)} & \ri \omega - v_F k\end{pmatrix} \right]. 
\ee 
The off-diagonal parts will enter only starting at the second order of the expansion of the log, thus $\alpha -\psi$ only enters with a prefactor of $J_\perp^2 \theta^4$, which is smaller than our accuracy and has to be neglected. 
Under this assumption, the angle $\alpha$ can be shifted to $\alpha-\psi$, thus eliminating one angular variable, as the Wess-Zumino term Eq. \eqref{WZ} also depends only on $\alpha+\psi$ to leading order in $\theta$ and $\alpha_\parallel$. 
It is easiest to eliminate $\alpha$ by bosonizing the modes coupled to the spins, and shifting \footnote{the same may be done in the EA case, as explained in Appendix \ref{sec:App_Shift}}
\be 
\label{trafo_ep} 
	\Theta_s \rightarrow \Theta_s - \sqrt 2 \alpha/4, \quad \Phi_c \rightarrow \Phi_c + \sqrt 2 \alpha/4. 
\ee 
The shift needs to be in both spin and charge sectors such that all charge conserving fermionic bilinears of the gapless sector remain unaffected. This is a consequence of the helical nature of the sectors and means that $\alpha$ will couple to both spin and charge sources: 
\be 
\label{source_helical_angle} 
	{\cal L}_{\rm source} \supset -\frac{h_c}{2 \pi} \partial_x \alpha - \frac{h_s}{2 \pi} \partial_x \alpha, 
\ee 
where we did not write the coupling of the sources to the fermions. 
Next, we integrate out the gapped helical sector. The ground state energy contribution from this is 
\be 
	\label{ground_state_energy_helical_fermions}
	\Delta E = - \frac{\xi_0}{2 \pi v_F} m^2 \ln (t/\vert m \vert) + {\cal O} (m^2), 
\ee 
where $m^2= \tfrac 12 \left (s \rho_s \cos \alpha_\parallel \cos \frac{\theta}{2} J_\perp \right )^2$. 
The ground state energy Eq. \eqref{ground_state_energy_helical_fermions} is minimized for $\alpha_\parallel = 0$ (we remind that $\theta \approx 0$). We expand $\Delta E$ to second order in $\alpha_\parallel$ and $\theta$ and obtain 
\be 
	%\Delta E \approx (s \rho_s)^2 \frac{\xi_0}{4 \pi v_F} \log (t/J_\perp) J_\perp^2 [\sin^2 (\theta/2) + \sin^2(\alpha_\parallel)/2] 
	\Delta E \approx -(s \rho_s)^2 \frac{\xi_0}{4 \pi v_F} \log (t/J_\perp) J_\perp^2 [(\theta/2)^2 + (\alpha_\parallel)^2]
\ee 
Thus, $\theta$ and $\alpha_\parallel$ are high-energy modes, which confirms the consistency of our approach in the EP phase. We can integrate out the fast variables and obtain
\be 
\label{LagrEP} 
	{\cal L}_{\rm ep} = R^\dagger_\downarrow G_R^{-1} R_\downarrow + L^\dagger_\uparrow G_L^{-1} L_\uparrow + \frac{1}{K_\alpha'}{\cal L}_{\rm TL} (\alpha, v_\alpha'), 
\ee 
where 
\be 
\label{velocitiesEP} 
	\frac{v_\alpha'}{v_F} = \frac{K_\alpha'}{4} = \frac{\xi_0 J_\perp}{2 \pi v_F} \sqrt{\log (t/J_\perp)} \ll 1,
\ee
and $G_{R/L}^{-1} = \partial_\tau \mp \ri v_F \partial_x$ is the inverse Green's function of free helical fermions. Upon bosonization, the gapless helical fermions become a helical Luttinger liquid: 
\be 
\label{LagrEP2} 
	{\cal L} = {\cal L}_{\rm TL} (\Phi^{\rm H1}, v_F) + \frac{1}{K_\alpha'} {\cal L}_{\rm TL} (\alpha, v_\alpha). 
\ee  
Thus, the low energy physics is described by two $U(1)$ Luttinger liquids, just as in the EA case. However, the Luttinger liquids are now helical modes and they differ from the EA case in the way they couple to external sources (c.f. Eq. \eqref{source_helical_angle}).

\subsection{The effects of electron interactions} 
\label{sec:interactions}
In the discussion of the EA and EP cases, we have neglected the effects of electron interactions. However, we used interactions to find the regime where the gap physics dominates Kondo physics. To fill this gap, we investigate the effects of interactions on the results of sections \ref{sec:easy_axis} and \ref{sec:easy_plane}. 
 
In the presence of interactions, $K_s$ and/or $K_c$ acquire values different from one. This changes the effect of the transformation Eq. \eqref{fermion-trafo} in the EA case. These transformations now induce terms of the form 
\be 
	{\cal L} \supset \frac{1}{2 \sqrt 2 \pi} \left( \frac{u_c}{K_c} \partial_x \alpha \partial_x \Phi_c - u_s K_s \partial_x \psi \partial_x \Theta_s \right). 
\ee 
Since all the fermions become massive, these terms may be dropped (c.f. discussion following Eq. \eqref{L_mixing}). The other effect of interactions is a renormalization of the gap $m$ (Eq. \eqref{GapsPM}). This is simply a renormalization of the parameters appearing in Eq. \eqref{L_sl_ea_1}, which we will neglect for now. 
 
In the EP case, the situation is different, because one helical branch remains gapless. If $K_s \neq 1/K_c$, the Luttinger parameter and the velocity of a helical sector (e.g. $R_\downarrow$ and $L_\uparrow$ as one sector) are changed to 
\be 
\label{K_u_interacting} 
	\tilde K = \sqrt{\frac{u_c K_c + \frac {u_s}{K_s}}{\frac{u_c}{K_c} + u_s K_s}}, \quad \tilde u = \frac{1}{2} \sqrt{u_c^2 + u_s^2 + u_c u_s K_c K_s + \frac{u_c u_s}{K_c K_s}}, 
\ee 
yielding the free part of the Lagrangian 
\be 
	{\cal L}_{h_i} = -\frac{\ri}{\pi} \partial_x \Theta_{h_i} \partial_\tau \Phi_{h_i} + \frac{1}{2\pi} \left (\tilde u \tilde K (\partial_x \Theta_{h_i})^2 + \frac{\tilde u}{\tilde K} (\partial_x \Phi_{h_i})^2 \right ). 
\ee 
Here, $\Phi_{h_i}$ is the bosonic field belonging to a given helical sector. 
The helical sectors $h_1$ (consisting of $R_\uparrow$ and $L_\downarrow$) and $h_2$ (consisting of $R_\downarrow$ and $L_\uparrow$) couple as  
\be 
	{\cal L}_{h-h} =  \frac{1}{2 \pi} \left \{ \left ( u_c K_c-\frac{u_s}{K_s} \right ) (\partial_x \Theta_{h_2} \partial_x \Theta_{h_1}) + \left ( \frac {u_c}{K_c}-u_s K_s \right ) (\partial_x \Phi_{h_2} \partial_x \Phi_{h_1}) \right \}
\ee 
The transformation Eq. \eqref{trafo_ep} thus adds to the Lagrangian the new part
\be 
\label{trafo_ep_interacting} 
	\delta {\cal L} =  \frac{1}{4 \pi}\left ( \frac {u_c}{K_c}- u_s K_s \right ) (\partial_x \Phi_{h_2} \partial_x \alpha) + {\cal O} (\partial \alpha \partial \Phi_{h_1}, \partial \alpha \partial \Theta_{h_1})  
\ee 
where $\Phi_{h_2}$ is the bosonic field belonging to the gapless (helical) fermionic modes. 
Dropping once more couplings of the derivative of the density of a gapped fermion (from the first helical sector) to gapless modes, the total low-energy Lagrangian ${\cal L}_{\rm ep}$ from  Eq. \eqref{LagrEP} is modified only by $\delta {\cal L}$ in Eq. \eqref{trafo_ep_interacting} \footnote{and a new effective Luttinger parameter and velocity, c.f. Eq. \eqref{K_u_interacting}}: 
\be 
\label{L_int_ep} 
	{\cal L}_{\rm ep}^{int} = {\cal L}_{h_2} + \frac{1}{K_\alpha'}{\cal L}_{\rm TL}(\alpha, v_\alpha') + \delta {\cal L}
\ee 
%\bea 
%\label{L_int_ep} 
%	{\cal L}_{\rm ep}^{int} &=& {\cal L}_{h_2} + \frac{1}{K_\alpha'}{\cal L}_{\rm TL}(\alpha, v_\alpha') + \delta {\cal L} \nonumber \\ 
%	                        &=& -\frac{\ri}{\pi} \partial_x \Theta_{h_2} \partial_\tau \Phi_{h_2} + \frac{1}{2\pi} \left [ \tilde u \tilde K (\partial_x \Theta_{h_2})^2 + \frac{\tilde u}{\tilde K} (\partial_x \Phi_{h_2})^2 \right ] + \frac{1}{2\pi}\left [ \frac{1}{K_\alpha'}(\partial_\tau \alpha)^2 + \frac{1}{K_\alpha'}(v_\alpha' \partial_x \alpha)^2 \right ] - \frac{1}{4 \pi}\left (\frac {u_c}{K_c}-u_s K_s \right ) \partial_x \alpha \partial_x \Phi_{h_2}. 
%\eea 
This expression can be analyzed by rediagonalizing it in field space. 
To do so, first integrate out $\Theta_{h_2}$. This yields 
\be 
\label{L_int_ep2} 
	{\cal L}_{\rm ep}^{int} = \frac{1}{2 \pi }\frac{1}{\tilde u \tilde K} (\partial_\tau \Phi_{h_2})^2 + \frac{1}{2\pi} \frac{\tilde u}{\tilde K} (\partial_x \Phi_{h_2})^2 + \frac{1}{2\pi} \left (\frac{1}{K_\alpha'}(\partial_\tau \alpha)^2 + \frac{1}{K_\alpha'}(v_\alpha' \partial_x \alpha)^2 \right ) + \frac{1}{4 \pi}\left (\frac {u_c}{K_c}-u_s K_s \right ) \partial_x \alpha \partial_x \Phi_{h_2}. 
\ee 
Next, we redefine the fields $\alpha$ and $\Phi_{h_2}$ such that the temporal derivatives have the same prefactor: 
\be 
	\alpha \rightarrow \sqrt{K_\alpha'} \alpha, \quad \Phi_{h_2} \rightarrow \sqrt{\tilde u \tilde K} \Phi_{h_2}  .
\ee 
This leads to 
\be 
\label{L_int_ep3} 
	{\cal L}_{\rm ep}^{int} = \frac{1}{2\pi}(\partial_\tau \Phi_{h_2})^2 + \frac{1}{2\pi} {\tilde u }^2(\partial_x \Phi_{h_2})^2 + \frac{1}{2\pi}(\partial_\tau \alpha)^2 + \frac{1}{2\pi}(v_\alpha' \partial_x \alpha)^2 + \delta \partial_x \alpha \partial_x \Phi_{h_2}, 
\ee 
where we have defined $\delta = \frac{1}{2 \pi} \sqrt{\tilde u \tilde K K_\alpha'}\left (\frac {u_c}{K_c} - u_s K_s \right )$. 
Diagonalizing this leads to two new gapless particles with dispersion 
\be 
	\omega^2 = \tfrac 12 \left( \tilde u^2 + v_\alpha^2 \pm \sqrt{(\tilde u^2 + v_\alpha^2)^2 + 4 \delta^2} \right) k^2. 
\ee 
Note that the remaining two degrees of freedom remain gapless. Interactions thus destroy the purely helical nature of low-energy excitations, but they cannot gap these exctiations.

\subsection{Suppression of forward scattering} 
We have seen that dominant backscattering leads to a vacuum structure where $\alpha_\parallel \approx 0$. The forward scattering terms however are proportional to $\sin \alpha_\parallel$, Eq. \eqref{CouplLagrSlowParamForward}. This confirms the suppression of their contribution once the gap is opened and examplifies our previous claim that Kondo physics and the gap physics are mutually exclusive.

\section{Density-density correlation functions and disorder} 
\label{sec:Density_correlations}
\subsection{Density-density correlation functions} 
We have shown that both the cases of EA and EP anistropy are described by two $U(1)$ Lutttinger liquids. However, the fields have different physical meaning as evinced by their coupling to external source. Their difference can be seen from various correlation functions. Let us at first consider the density-density correlation function 
\be 
	{\cal C} = \left \langle \rho_c (1) \rho_c (2) \right \rangle = \frac{\delta^2 \log Z[h_c]}{\delta h_c (1) \delta h_c (2)} \vert_{h_c =0}, 
\ee 
where $\rho_c$ is the electron density and $Z[h_c]$ is the generating functional in the presence of the source $h_c$. In general, there are several contributions to ${\cal C}$, including those from gapped and gapless excitations. Even if the fermionic modes become gapped, there still is a contribution from collective electron and spin modes to long range density-density correlation functions. This can be seen from the fact that some low energy degrees of freedom (EA: $\alpha$; EP $\alpha$ and one helical fermion) couple to $h_c$. In Fourier space, the correlation functions are
\bea 
	{\cal C}_{\rm ea} (\omega, q) &=& \left (\frac{q}{\pi}\right )^2 \left \langle \alpha^* \alpha \right \rangle, \\ 
	{\cal C}_{\rm ep} (\omega, q) &=& \left (\frac{q}{\pi}\right )^2 \left ( \left \langle \Phi_H^* \Phi_H \right \rangle + \left \langle \alpha^* \alpha \right \rangle /4 \right ) .
\eea 
Using the corresponding low energy effective actions Eq. \eqref{LagrEA} and Eq. \eqref{LagrEP}, this yields 
\bea 
\label{Dens-Dens-correlator1}
	{\cal C}_{\rm ea} (\omega, q) &=& \frac{q^2 K_\alpha v_\alpha^2 \xi_0^{-2}}{\pi (\omega^2 + (v_\alpha q)^2)}, \\ 
\label{Dens-Dens-correlator2}
	{\cal C}_{\rm ep} (\omega, q) &=& \frac{q^2}{\pi} \left ( \frac{ v_F^2 \xi_0^{-2}}{\omega^2 + (v_F q)^2}+\frac {1}{4} \frac{K_\alpha' {v_\alpha'}^2 \xi_0^{-2}}{\omega^2 + (v_\alpha' q)^2} \right ). 
\eea 
Equations \eqref{Dens-Dens-correlator1} and \eqref{Dens-Dens-correlator2} correspond to ideal metallic transport. The small Luttinger parameter of the bosonic modes ($K_\alpha, \hspace{1 pt} K_\alpha' \ll 1$) reflects the coupling of the spin waves to the gapped fermions and leads to a reduced Drude weight \cite{AAY}.

\subsection{The role of potential disorder} 
 
Let us investigate how potential disorder affects charge transport. We add a weak random potential 
\be 
	\label{eq:disorder_pot} 
	V_{\rm dis} = g(x) \Psi^\dagger ({I} \otimes {\tau^+}) \Psi + H.c., 
\ee 
where $g(x)$ is the smooth $2 k_F$ component of the scalar random potential. Note that we have dropped quickly oscillating modes, just as for the spin impurities. If the disorder itself is distributed according to the Gaussion orthogonal ensemble (GOE), then its $2 k_F$ component has a Gaussian unitary distribution. Thus the function $g$ is drawn from a Gaussian unitary ensemble (GUE). We use $\left \langle g(x) \right \rangle = 0$ and $\left \langle g^*(x) g(y) \right \rangle = 2 {\cal D} \delta(x-y)$.
We assume that the potential disorder is sufficiently weak, such that it does not influence the high energy physics. The precise meaning of this statement will be specified later. 
 
As first step, we integrate the disorder exactly by using the replica trick. Upon disorder-averaging we obtain 
\be 
	{S}_{\rm dis} = \sum_{i,j} \int dx \int d\{\tau_{1,2} \} {\cal D} \left[ \left (R^\dagger_{\uparrow i} L_{\uparrow i} + (\uparrow \leftrightarrow \downarrow) \right )(x, \tau_1) \left (L^\dagger_{\uparrow j} R_{\uparrow j} + (\uparrow \leftrightarrow \downarrow) \right )(x, \tau_2) \right], 
\ee 
where $i,j$ are replica indices. The remainder of the action is diagonal in replica space.  
 
To understand the effect of ${S}_{\rm dis}$ on transport we now have to integrate out the massive modes. Recall that this involves first a shift of the fermionic fields (Eq. \eqref{fermion-trafo}) \footnote{In EP, the shift leads to the same result after absorbing $\psi$ in $\alpha$}: 
\be 
	{S}_{\rm dis} = \int dx \int d\{\tau_{1,2} \} {\cal D} \left[ \left (R^\dagger_{\uparrow i} L_{\uparrow i} \re^{\ri \alpha_i} + (\uparrow \leftrightarrow \downarrow) \right )(x, \tau_1) \left (L^\dagger_{\uparrow j} R_{\uparrow j} \re^{-\ri \alpha_j} + (\uparrow \leftrightarrow \downarrow) \right )(x, \tau_2) \right], 
\ee 
where the gapped and gapless modes now are cleanly separated in the rest of the action (with our accuracy). Thus, it is easy to integrate out the gapped modes. We treat $S_{\rm dis}$ perturbatively, obtaining an expansion in the parameter $\frac{\cal D }{v_F m} \ll 1$ (weak disorder).  
 
In the EA case, all fermions are gapped and the only gapless mode appearing in ${\cal L}_{\rm dis}$ is the charge mode $\alpha$. In the EP case only the fermions with a given helicity (e.g. $R_{\uparrow}$ and $L_{\downarrow}$) become gapped and the disorder mixes the two helical Luttinger liquids ($\alpha$ and the fermions of the non-gapped helicity).  
It is convenient to treat EA and EP separately. 
 
\subsubsection{Easy axis} 
We start with the EA case, and put $J_\perp = 0$. For transparency, we choose the fermionic spin-dependent mass $m_{\rm ea}(\uparrow/\downarrow) = \pm m$. 
The matrix Green's function for the fermions with a given spin reads: 
\be 
\label{GFm} 
  \hat{G}_m(\sigma) = \left( (G^{(0)}_R)^{-1} (G^{(0)}_L)^{-1} - m_{\rm ea}(\sigma)^2  \right)^{-1} 
      \left( 
         \begin{array}{cc} 
                \left( G^{(0)}_L \right)^{-1}  & -  m_{\rm ea}(\sigma)          \\ 
                    -  m_{\rm ea}(\sigma)     &  \left( G^{(0)}_R \right)^{-1} 
         \end{array} 
      \right) ; 
\ee 
where $ G^{(0)}_{R,L} $ are the Green's functions of free chiral particles. It is important 
that $ \hat{G}_m $ is short ranged and it decays beyond the time scale $ 1/m_{\rm ea} $ 
(or beyond the coherence length $ \xi_{\rm ea} \equiv v_f/m_{\rm ea} $).  
This implies in particular that two slow operators connected by a massive propagator form a single local operator on length- and timescales large compared to the inverse gap. 
 
\begin{figure}[h] 
   \includegraphics{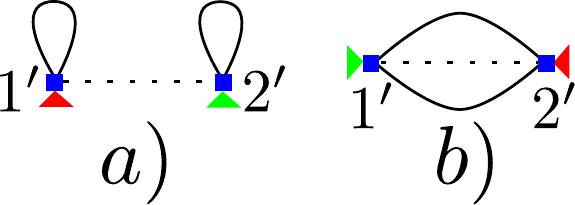} 
   \caption{\label{EA-d1} 
        First order diagrams $ \, O({\cal D}^1) \, $ for the EA phase. Red (green) triangles 
        denote $ \, \re^{\ri \alpha/2} \, $ ($ \, \re^{-\ri \alpha/2} \, $) with arguments of either the 1st 
        or the 2nd vertex; dashed lines are the disorder correlation functions, solid lines stand 
        for Green's functions of the massive fermions. 
                 } 
\end{figure}

Leading terms are given by $ \left \langle S_{\rm dis} \right \rangle_m $ 
where brackets mean that the massive fermions are integrated out. The corresponding 
diagrams are shown in Fig.\ref{EA-d1}. It is easy to check that the diagrams from Fig.\ref{EA-d1}-a 
cancel out after summation over spin indices because $ \, m_{\rm ea} (\uparrow) = - m_{\rm ea} 
(\downarrow) $. The diagrams from Fig.\ref{EA-d1}-b are trivial since $ \hat{G}_m $ is diagonal 
in the replica space and the spin phase $ \alpha $ is smooth on the scale $ 1/m_{\rm ea} $; therefore, 
\be 
  \label{Canc-a} 
  \re^{\ri \alpha[{\bf 1}]} \re^{-\ri \alpha[{\bf 2}]} \simeq \re^{\ri \alpha[{\bf 1}]-\ri \alpha[{\bf 1}]} = 1 \, , 
\ee 
with some small gradient corrections which are unable to yield pinning. Here we denoted $\alpha[j] := \alpha(x_1, \tau_1)$.
 
\begin{figure}[h] 
   \includegraphics{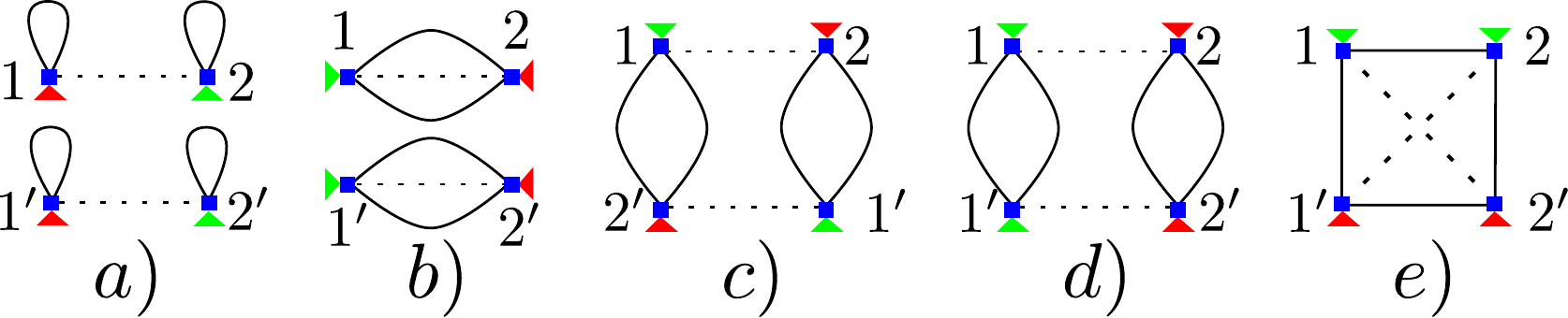} 
   \caption{\label{EA-d2} 
	A relevant subset of the EA diagrams. Notations are explained in the caption of Fig. \ref{EA-d1}. 
	a) and b) Class (ii), diconnected contributions. c) Class (iii), red and green triangles are merged through a massive propagator. d) Class (iv), we omit the diagram with crossed disorder lines. e) Class (v), we omit the diagram with non-crossed disorder lines. Note that green and red triangles are connected by a massive propagator. 
                 } 
\end{figure}

Sub-leading terms of the order of $ \, \frac{{\cal D}^2}{v_F m} \, $ are given by $ \left \langle S_{\rm dis} 
S_{\rm dis} \right \rangle $. To be explicit, we need to compute 
\bea 
	\left \langle {S}_{\rm dis} {S}_{\rm dis} \right \rangle_{\rm EA}= {\cal D}^2 \Big \langle \int d\{x, x'; \tau_{1,2}, \tau_{1,2}' \} & \left[  \left (R^\dagger_{\uparrow i} L_{\uparrow i} \re^{\ri \alpha_i} + (\uparrow \leftrightarrow \downarrow) \right )(x, \tau_1) \left (L^\dagger_{\uparrow j} R_{\uparrow j} \re^{-\ri \alpha_j} + (\uparrow \leftrightarrow \downarrow) \right )(x, \tau_2) \right] \nonumber \\ 
	 & \left[  \left (R_{\uparrow k}^\dagger L_{\uparrow k} \re^{\ri \alpha_k} + (\uparrow \leftrightarrow \downarrow) \right )(x', \tau_1') \left (L^\dagger_{\uparrow l} R_{\uparrow l} \re^{-\ri \alpha_l} + (\uparrow \leftrightarrow \downarrow) \right )(x', \tau_2') \right] \Big \rangle_{\rm EA}. 
\eea 
In order to pin the CDW (the field $\alpha$), an operator evaluating $\alpha$ at different times (i.e. times further apart than $1/m_{\rm ea}$) has to survive. 
The correlation function $\left \langle {S}_{\rm dis} {S}_{\rm dis} \right \rangle_{\rm EA}$ contains various possible contractions, most of which are unable to generate pinning: 
\begin{enumerate}[(i)] 
	\item Contractions involving two fermionic creation or annihilation operators: They vanish due to the structure of the fermionic Green's function, which does not allow for propagation of Cooper pairs.  
	\item Contractions which simplify to two copies of the first order contribution (c.f. Fig. \ref{EA-d2} a, b): They do not generate backscattering, as shown above. 
	\item Contractions of fermions at $(x, \tau_1)$ with fermions at $(x', \tau_2')$ and of fermions at $(x, \tau_2)$ with fermions at $(x', \tau_1')$, with no contractions between $(x, \tau_1)$ and $(x', \tau_1')$ (Fig. \ref{EA-d2} c): In these contractions - due to the short range nature of the fermions' Green's functions - $\re^{\ri \alpha}$ fuses with $\re^{-\ri \alpha}$ at the same position and time (at an accuracy of $1/m$), and thus generate only derivatives of $\alpha$, which are unable to pin the CDW. 
	\item Contractions of fermions at $(x, \tau_1)$ with fermions at $(x', \tau_1')$ and of fermions at $(x, \tau_2)$ with fermions at $(x', \tau_2')$, with no contractions between $(x, \tau_1)$ and $(x', \tau_2')$ (Fig. \ref{EA-d2} d): These contractions all give the same result and are able to generate pinning. 
	\item Contractions between all positions and times (Fig. \ref{EA-d2} e): This sets all positions and times (and replica indices) of the CDW equal to each other (with accuracy $1/m$), such that again only derivatives of the field $\alpha$ survive. 
\end{enumerate} 
 
We calculate only one typical diagram which survives after all summations  
and is able to generate pinning [type (iv)]. An example of such a diagram is shown in Fig.\ref{EA-d2}d. All other diagrams of class (iv) yield identical results. The sign of the mass does not matter as there is an even number of propagators for each species. 
 
Neglecting unimportant numerical factors, the analytical expression for the diagram 
from Fig. \ref{EA-d2}d reads as: 
\be 
\label{D2-EA} 
  D^{(2)}_{\rm ea} \propto 
  {\cal D}^2 \sum_{i,j} \int {\rm d} \{ x, x' ; \tau_{1,2}, \tau_{1,2}' \} 
  \re^{2 \ri \left( \alpha_i[{\bf 1}] -  \alpha_j[{\bf 2}] \right)} 
  \left[ \hat{G}_m({\bf 1,1'}) \right]_{1,2} \left[ \hat{G}_m({\bf 1',1}) \right]_{1,2} 
  \left[ \hat{G}_m({\bf 2,2'}) \right]_{1,2} \left[ \hat{G}_m({\bf 2',2}) \right]_{1,2} \, . 
\ee 
Here, we have taken into account that the diagonal structure of $ \hat{G}_m $ results 
in $ i = k; j = l $ 
and fused together slow spin phases, for instance: $ \alpha[{\bf 1}] + \alpha[{\bf 1'}] \simeq 
2 \alpha[{\bf 1}] $. Now we note that $ \hat{G}_m({\bf 1,1'}) = \hat{G}_m({\bf 1 - 1'}) $ 
and integrate over all primed variables: 
\be 
\label{EA-SG} 
  D^{(2)}_{\rm ea} \propto 
  \frac{ \tilde{{\cal D}}_0 }{\xi_{\rm ea}^2} 
  \sum_{i,j} 
  \int {\rm d} \{ x; \tau_{1,2} \} \re^{\ri \left( \alpha_i[{\bf 1}] -  \alpha_j[{\bf 2}] \right)} \, ; \quad 
  \tilde{ {\cal D} }_0 \equiv {\cal D} \left( \frac{{\cal D}}{v_F m_{\rm ea}} \right) . 
\ee 
The structure of Eq. \eqref{EA-SG} corresponds to the non-local Sine-Gordon model 
which appears in the theory of the usual disordered TLL \cite{Giamarchi}. The effective disorder strength 
$  \tilde{ {\cal D} } $ is renormalized and obeys the well-known RG equation \cite{GiaSchulz}: 
\be 
\label{RG-EA} 
  {\mbox EA}: \quad 
  \partial_{\log} \log( \tilde{ {\cal D} } ) = 3 - 2 K_\alpha \simeq 3 \, , \quad 
  \tilde{ {\cal D} }(\xi_{\rm ea}) = \tilde{{\cal D}}_0 \, ; 
\ee 
the second equality of Eq.(\ref{RG-EA}) has been obtained by using Eq.(\ref{Compr-a}). 
 
Note that the effective strength of the disorder is suppressed compared to free fermions by an additional factor of ${\cal D}/(v_F m)$. 
However, the operator is more relevant than for free fermions, as $K_\alpha^{\rm (EA)} \ll 1$.

\subsubsection{Easy Plane} 
\begin{figure}[h] 
   \includegraphics{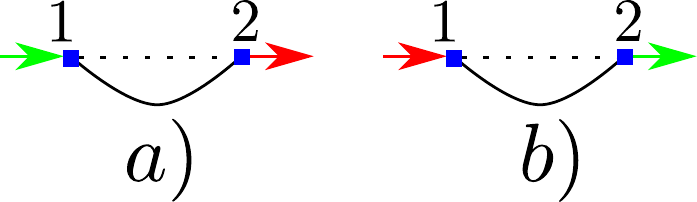} 
   \caption{\label{EP-d1} 
        Two typical examples of first order diagrams $ \, O({\cal D}^1) \, $ for the EP phase. 
        Red (green) arrows 
        denote the product of smooth fields $L$, $R$ with $ \, \re^{\ri \alpha/2} \, $ ($ \, \re^{-\ri \alpha/2} \, $). The smooth fields $L$, $R$ are taken from the non-gapped helical sector. 
                 } 
\end{figure}

Let us now turn to the EP case.  
We start again from the leading diagrams generated by $ \left \langle S_{\rm dis} \right \rangle $. 
The principal difference of the EP phase from the EA one is that the matrix Green's 
function, Eq.(\ref{GFm}), corresponds now to the massive fermions with a given helicity. 
This changes the structure of the first order diagram, see Fig.\ref{EP-d1}. All these 
diagrams correspond to forward-scattering of the massless helical fermions and 
they contain only small gradients of the phase $ \, \alpha $, cf. Eq.(\ref{Canc-a}) and 
its explanation. Thus, the leading diagrams are trivial and they cannot yield localization, 
the sub-leading diagrams must be considered. 
 
\begin{figure}[h] 
   \includegraphics{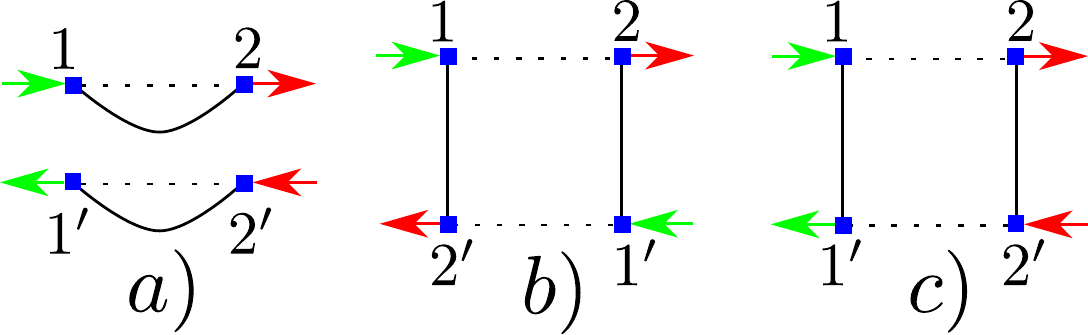} 
   \caption{\label{EP-d2} 
	A relevant subset of the EP diagrams. Notations are explained in the caption of Fig. \ref{EP-d1}. 
	a) Class (ii), diconnected contributions. b) Class (iii), red and green arrows are merged through a massive propagator. We omit the diagram with crossed disorder lines. Note that green and red arrows are connected by a massive propagator. c) Class (iv), we omit the diagram with crossed disorder lines. 
                 } 
\end{figure}

There are several categories of sub-leading diagrams: 
\begin{enumerate}[(i)] 
	\item Contractions involving two creation or annihilation operators: They are identically zero. 
	\item Contractions which correspond to two copies of the leading diagrams (Fig. \ref{EP-d2} a): They do not lead to backscattering and cannot pin the charge transport. 
	\item Contractions of fermions at $(x, \tau_1)$ with fermions at $(x', \tau_2')$ and of fermions at $(x, \tau_2)$ with fermions at $(x', \tau_1')$, with no contractions between $(x, \tau_1)$ and $(x', \tau_1')$ (the second part - excluding certain contractions - is trivial, as there is only one massive fermion at each vertex) (Fig. \ref{EP-d2} b): These contractions - due to the short range nature of the fermions' Green's function - combine $\re^{\ri \alpha}$ with $\re^{-\ri \alpha}$ at the same position and time (at an accuracy of $1/m$), and thus generate only derivatives of $\alpha$, which are unable to pin the CDW. 
	\item Contractions of fermions at $(x, \tau_1)$ with fermions at $(x', \tau_1')$ and of fermions at $(x, \tau_2)$ with fermions at $(x', \tau_2')$, with no contractions between $(x, \tau_1)$ and $(x', \tau_2')$ (the second condition is again trivially satisifed) (Fig. \ref{EP-d2} c). These contractions all give the same result and are able to generate pinning. 
\end{enumerate} 
 
The only relevant diagrams are those of class (iv), which all yield the same result. We will compute one of these diagrams (Fig. \ref{EP-d2}c). 
Neglecting unimportant numerical factors, the analytical expression for the diagram 
from Fig.\ref{EP-d2}c reads as: 
\be 
  D^{(2)}_{\rm ep} \propto 
  {\cal D}^2 \sum_{i,j} \int {\rm d} \{ x, x' ; \tau_{1,2}, \tau_{1,2}' \} 
  \re^{\ri \left( \alpha_i[{\bf 1}] -  \alpha_j[{\bf 2}] \right)} \, 
  L^\dagger_{\downarrow j}[ {\bf 2} ] R^\dagger_{\uparrow i}[ {\bf 1} ] \, L_{\downarrow i}[ {\bf 1} ] R_{\uparrow j}[ {\bf 2} ] \, 
  \left[ \hat{G}_m({\bf 1,1'}) \right]_{1,2} 
  \left[ \hat{G}_m({\bf 2,2'}) \right]_{1,2} \, ; 
\ee 
see explanations after Eq.(\ref{D2-EA}) and note the $ m $ must be substituted for $ m_{\rm ea} (\sigma) $ 
in $ \hat{G}_m $.  Calculating integrals over all primed variables, we find: 
\be 
\label{EP-SG} 
  D^{(2)}_{\rm ep} \propto 
  \bar{{\cal D}}_0 
  \sum_{i,j} \int {\rm d} \{ x ; \tau_{1,2} \} 
  \re^{\ri \left( \alpha_i[{\bf 1}] -  \alpha_j[{\bf 2}] \right)} \, 
  L^\dagger_{\downarrow j}[ {\bf 2} ] R^\dagger_{\uparrow i}[ {\bf 1} ] \, L_{\downarrow i}[ {\bf 1} ] R_{\uparrow j}[ {\bf 2} ] \, , \quad 
  \bar{ {\cal D} }_0 \equiv {\cal D} \left( \frac{{\cal D}}{v_F m} \right) . 
\ee 
This equation also can be reduced to the form of Eq.(\ref{EA-SG}) if remaining fermions are bosonized 
and we explicitly single out new charge- and spin- density waves. However, the RG equation for 
$ \bar{ {\cal D} } $ can be obtained without such a complicated procedure with the help of the 
power counting. Firstly we note that the scaling dimension of each back-scattering term in Eq.(\ref{EP-SG}), 
$ L^\dagger R $ and $ R^\dagger L $, is $1$. The anomalous dimension of each exponential, $ \re^{\pm \ri 
\alpha} $, is $ K_\alpha' \ll 1 $. The normal dimension in Eq.(\ref{EP-SG}) is $3$ which comes from three-fold 
integral. Combining these dimensions together and neglecting small $ K_\alpha'  $, we find 
\be 
\label{RG-EP} 
  {\mbox EP}: \quad 
  \partial_{\log} \log( \bar{ {\cal D} } ) = 3 - 2 \times 1 + O(K_\alpha) \simeq 1 \, ; \quad 
  \bar{ {\cal D} }(\xi_{\rm ep}) = \bar{{\cal D}}_0 \, , \ \xi_{\rm ep} = v_F / m \, . 
\ee 
Note that while the scaling of the disorder strength is the same as for free fermions, but the effective strength (the starting value of the flow) is reduced parametrically by a factor of ${\cal D}/(v_F m) \ll 1$. 
 
\subsubsection{Localization Radius} 
We now can find the localization radius for both phases, EA and EP. 
The solution of the RG equations, Eqs.(\ref{RG-EA},\ref{RG-EP}), reads as 
\be 
  \tilde{\cal D}(x) = \tilde{\cal D}_0 \left( \frac{x}{\xi_{\rm ea}} \right)^3, \quad 
  \bar{\cal D}(x) = \bar{\cal D}_0 \frac{x}{\xi_{\rm ep}} \, ; 
\ee 
with $ \xi_{\rm ep} = v_F / m $. The localization radius is defined as a scale on 
which the renormalized disorder becomes of the order of the cut-off: 
\be 
\label{LlocEq} 
   \tilde{\cal D}\left(L^{\rm (loc)}_{\rm ea}\right) = 
      K_\alpha v_\alpha^2 / \xi_{\rm ea} \sim K_\alpha^3 v_F^2 / \xi_{\rm ea} \, ; \quad 
   \bar{\cal D}\left(L^{\rm (loc)}_{\rm ep}\right)  = v_F^2 / \xi_{\rm ep} \, . 
\ee 
The additional small factor $ K_\alpha $ in the equation for $ L^{\rm (loc)}_{\rm ea} $ can be justified 
with the help of the standard optimization procedure \cite{Giamarchi} where $ L^{\rm (loc)} \, $ is 
defined as a spatial scale on which the typical potential energy of the disorder becomes equal to 
the energy governed by the term $ \, \propto (\partial_x \alpha)^2 \, $ in the Lagrangian 
$ {\cal L}_{\rm ea} $, Eq.(\ref{LagrEA}). 
 
Definitions Eq.(\ref{LlocEq}) result in 
\be 
  \label{eq:Localization_radii} 
  L^{\rm (loc)}_{\rm ea} \sim  \xi_{\rm ea} K_\alpha \left(  \frac{v_F^2}{ \xi_{\rm ea} \tilde{\cal D}_0 } \right)^{1/3} \!\!\! \sim 
                                               \xi_{\rm ea} K_\alpha \left( \frac{v_F m_{\rm ea}}{ {\cal D} } \right)^{2/3} \!\!\! ; \quad 
  L^{\rm (loc)}_{\rm ep} \sim \frac{v_F^2}{\bar{\cal {D}}_0} \sim 
                                              \xi_{\rm ep} \left( \frac{v_F m}{{\cal D}} \right)^2 \, . 
\ee 
Assuming $ \xi_{\rm ea} \sim \xi_{\rm ep}$ and $ m_{\rm ea} \sim m $, we obtain 
\be 
  \frac{L^{\rm (loc)}_{\rm ea}}{L^{\rm (loc)}_{\rm ep}} \sim K_\alpha \left( \frac{{\cal D}}{v_F m} \right)^{4/3} \ll 1 \, . 
\ee 
This demonstrates that the strong suppression of localization can occur in the EP phase where the 
helical symmetry is broken. 
 
We note that the scaling exponent of $ \bar{\cal D}(x) $ is the same as in 
the case of non-interacting 1d fermions but suppression of localization in the EP 
phase is reflected by the additional large factor $  v_F m / {\cal D} $ in the expression 
for the localization radius $ L^{\rm (loc)}_{\rm ep} $. 
We further note that unlike for free fermions our flow starts at the correlation length $v_F/m$, not at the lattice constant $\xi_0$. However, for characteristic length scales $\xi_0 < l < \xi_{{\rm ea}/{\rm ep}}$, the mass is not relevant and the flow of our system mimics that of free fermions in the absence spinful impurities. The flow only begins to differ at $l \approx \xi_{{\rm ea}/{\rm ep}}$, such that we should compare to free fermions with a cutoff $\xi_{{\rm ea}/{\rm ep}}$.

\subsubsection{Alternative approach to disorder} 
 
In this section we present an alternative approach which confirms the previous results on disorder. 
The main idea is to integrate out the massive modes before averaging over disorder.
We will focus on the main steps and neglect unimportant prefactors. 
Let us start again at Eq. \eqref{eq:disorder_pot}. 
In the EA case, we perform a shift $\Phi_c \rightarrow \Phi_c - \alpha/\sqrt 2$. This shift leads to  
\be 
	V_{\rm dis}^{\rm ea} = g(x) \re^{\ri \alpha} \Psi^\dagger ({I} \otimes {\tau^+}) \Psi + H.c., 
\ee 
such that the field $\alpha$ couples to the potential disorder. 
Let us integrate out the massive fermions. The leading term (in powers of the disorder) in the Lagrangian is then 
\be 
	L_{\rm dis} \sim \frac{1}{\xi_0} \int \rd x \Big[g_{\rm eff}(x) \re^{\ri 2\alpha} + H.c.\Big], \label{potdisorder} 
\ee 
where we introduced the non-Gaussian effective disorder 
\be 
	g_{\rm eff}(x) \sim \frac{1}{2 v_F}\int \rd y g(x+y/2)g(x-y/2)\re^{-m|y|/v_F}; \label{effect} 
\ee 
the exponential stems from real space Green's function of fermions with mass $m$ \footnote{Note that Eq. \eqref{potdisorder} corresponds to Fig. \ref{EA-d2} c,d: The fermionic lines are contracted to a single point and the two disorder lines are merged into one line corresponding to $g_{\rm eff}$}. Eq. \eqref{effect} is valid for large distances $y \gg v_F/m$.

In the EP case, before integrating out the massive fermions, we shift their phase $\Phi_c$ by $\sqrt 2 \alpha/4$: 
\be 
	V_{\rm dis}^{ep} = \int \rd x \Big[g(x) \re^{\ri \alpha /2} R^\dagger_\uparrow L_\uparrow + g(x) \re^{\ri \alpha /2} R^\dagger_\downarrow L_\downarrow  + H.c.\Big] .
\ee 
Each term describes a coupling of a gapped fermion from the first helical sector with a gapless one from the second helical sector and with a low-energy angle $\alpha$.  
Upon integrating out the gapped fermions,
%(whose action is quadratic in the fermionic fields in our approximation),
the disorder generates the following contribution to the low energy effective Lagrangian: 
\be 
\label{potdisorder_ep}
	{\cal L}_{\rm dis}^{\rm (H2)} \supset  \int \rd x \Big[g_{\rm eff}(x) R^\dagger_\downarrow \re^{\ri \alpha} L_\uparrow + H.c.\Big], 
\ee 
where $g_{\rm eff}(x)$ is of the form of Eq. \eqref{effect}. \footnote{Eq. \eqref{potdisorder_ep} corresponds to contracting the internal fermion lines in Fig. \ref{EP-d2} b,c, and then merging the two disorder lines into a single lines described by $g_{\rm eff}$.}
 
Thus, both in EA and EP, we obtain gapless particles 
%(EA: CDW and SDW; EP: Helical fermions and a helical Luttinger liquid) 
coupled to an effective disorder.

To order $\frac{{\cal D}}{v_F m}$, only the first and second moment of the distribution function of $g_{\rm eff}$ contribute (see Appendix \ref{Appendix_effective_disorder}). This is equivalent to the statement that the non-Gaussianities of the distribution of $g_{\rm eff}$ are irrelevant in our approximation.
 
%We can remedy this, as it is usually done, by replacing $g_{eff}$ with the scattering amplitude. If the bare disorder potential is small in comparison with the fermion gap $m$ this will not change the qualitative picture. Namely, we can estimate the disorder effects by averaging over the disorder in the replicated action, as it was done in \cite{Giamarchi}. Then the part of the action associated with the disorder is 

The leading order contributions of the effective disorder to the localization may then be estimated similarly to the diagrammatic approach. Upon integrating over the disorder (and assuming it's a Gaussian distribution), we obtain
\bea 
	S_{dis} \sim \sum_{i,j} \int \rd\tau\rd\tau'\rd x \frac{{\cal D}^2}{v_F m} {\cal O}_i(x,\tau){\cal O}^\dagger_j(x,\tau'), 
\eea 
where the operator $\cal O$ is given by
\bea
	EA: \, {\cal O}_i(x,\tau) = \frac{1}{\xi_0} \re^{\ri 2\alpha_i(x,\tau)}, \\
	EP: \, {\cal O}_i(x,\tau) = \re^{\ri\alpha_i}R^\dagger_{i,\downarrow}L_{i,\uparrow}.
\eea 
%We can estimate the pinning energy by the power counting: 
%\be 
%	{\rm EA:} \, \epsilon_0\sim \gamma^{1/(3- 2K)}, ~~ {\rm EP:} \, \epsilon_0 \sim \gamma^{1/(1-K)}, 
%\ee 
%where $\gamma = \frac{{\cal D}^2}{v_F m}$.
This yields the same scaling and, thus, the same localization radius Eq. \eqref{eq:Localization_radii} as in the diagrammatic approach. 

%The leading order contributions of the effective disorder to the localization may then be estimated as follows: We notice, that the action is of the form 
%\bea 
%	S_{dis} = \sum_{a,b=1}^r\int \rd\tau\rd\tau'\rd x g_{\rm eff}(x) {\cal O}_a(x,\tau){\cal O}^+_b(x,\tau'), 
%\eea 
%where in the easy axis case  
%\be 
%	{\cal O}_a(x,\tau) = \frac{1}{\xi_0} \re^{\ri 2\alpha_a(x,\tau)}, 
%\ee 
%and in the easy plane one  
%%where in the easy plane case  
%\be 
%{\cal O}_a(x,\tau) = \re^{\ri\alpha_a}R^+_{a,\downarrow}L_{a,\uparrow} 
%\ee 
%and $g_{\rm eff}$ is distributed according to the GUE, with
%\be
%	\left \langle g_{\rm eff}(x) g_{\rm eff}^*(y) \right \rangle \sim \frac{{\cal D}^2}{v_F m}.
%\ee
%%\bea 
%%	\gamma \sim 2 g^2\Big[\int \rd x G_{RL}^2(\omega=0,x)\Big] = \frac{g^2}{2m} 
%%\eea 
%with $m$ being the mass of the gapped fermions. We can estimate the pinning energy by the power counting: 
%\be 
%	{\rm EA:} \, \epsilon_0\sim \gamma^{1/(3- 2K)}, ~~ {\rm EP:} \, \epsilon_0 \sim \gamma^{1/(1-K)} . 
%\ee 
%This yields the same scaling and thus the same localization radius Eq. \eqref{eq:Localization_radii} as in the diagrammatic approach. 
The advantage of this approach is that the order of approximations follows the ordering of the relevant energy scales. We first eliminate the highest energy ($m$) and only then approach the much smaller pinning energy. The price is the non-Gaussianity of the effective disorder. However, since higher moments of the effective disorder are suppressed by additional factors of $\frac{\cal D}{v_F m}$, the non-Gaussianities only enter in higher orders that we do not consider here.

\section{Spin correlation functions and order parameter} 
\label{sec:spins} 
\begin{figure} 
\includegraphics[scale=.8]{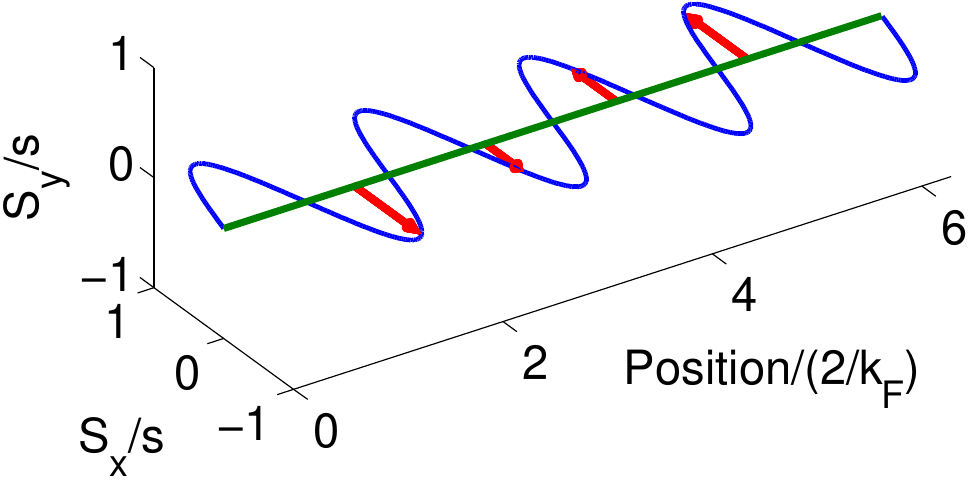} 
\quad \quad 
\includegraphics[scale=.8]{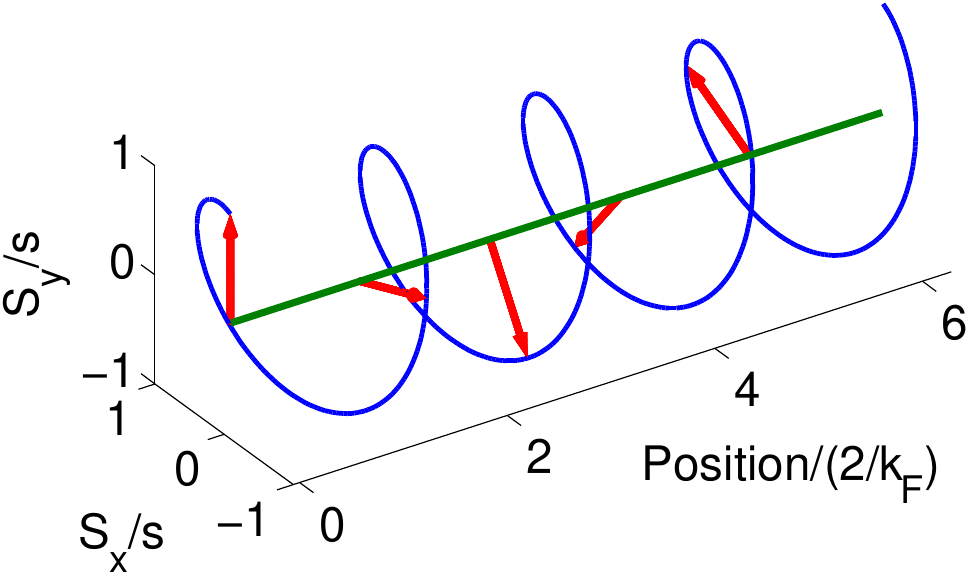} 
\caption{(color online) A travelling spin wave in the EA (left) and EP (right) setup. Since $S^x$ and $S^y$ in the EA case are uncorrelated to leading order, we only show one contribution.} 
\label{fig:spin-wave} 
\end{figure} 
Let us consider the spin correlators $\left \langle S^a (1) S^b (2) \right \rangle$ and see which correlation function reflects the broken $\mathbb Z_2$ symmetry. % and, thus, might be used to define an order parameter. 
 
Before computing the correlators, we note the following: The low energy physics of both phases is captured by two uncorrelated $U(1)$ Luttinger liquids and by a set of fast angles. The slow component of the spins (in the basis where $S_{slow} \parallel e_3'$) depends on the angles via 
\begin{subequations}
\label{eq:spins_general} 
\bea 
	S^x/s & = & -\cos \alpha_\parallel \cos \alpha_\perp \cos \theta \cos \psi + \cos \alpha_\parallel \sin \alpha_\perp \sin \psi + \sin \alpha_\parallel \sin \theta \cos \psi ;  \\ 
	S^y/s & = & -\cos \alpha_\parallel \cos \alpha_\perp \cos \theta \sin \psi - \cos \alpha_\parallel \sin \alpha_\perp \cos \psi + \sin \alpha_\parallel \sin \theta \sin \psi  ; \\ 
	S^z/s & = &  \cos \alpha_\parallel \cos \alpha_\perp \sin \theta                                                               + \sin \alpha_\parallel \cos \theta          . 
\eea 
\end{subequations}
The effective low energy physics is generated at $\alpha_\parallel \approx 0$. Therefore, Eq. \eqref{eq:spins_general} simplifies to  
\begin{subequations}
\bea 
	S^x/s & = & -\cos \alpha_\perp \cos \theta \cos \psi + \sin \alpha_\perp \sin \psi ; \\ 
	S^y/s & = & -\cos \alpha_\perp \cos \theta \sin \psi - \sin \alpha_\perp \cos \psi ; \\ 
	S^z/s & = &  \cos \alpha_\perp \sin \theta                                        ; 
\eea 
\end{subequations}
where we neglect fast fluctuations of $\alpha_\parallel$ around its ground state value. 
We will also need the correlation functions (for large distances) in a Luttinger liquid described by the field $\rho$ with Luttinger parameter $K$ and velocity $v$:
\begin{subequations}
\label{basic_correlation_functions_LL}
\bea
	\left \langle \sin (\rho(x_1) \pm \rho(x_2)) \right \rangle = 0; \quad \left \langle \cos(\rho(x_1) + \rho(x_2)) \right \rangle = 0; \quad \left \langle \sin (\rho(x_1)) \cos (\rho(x_2)) \right \rangle = 0;  \\
	\left \langle \sin (\rho(x_1)) \sin (\rho(x_2)) \right \rangle = \left \langle \cos (\rho(x_1)) \cos (\rho(x_2)) \right \rangle = \tfrac 12 \left \langle \cos (\rho(x_1) - \rho(x_2)) \right \rangle = \frac{\xi_0^{K/2}}{[(v \tau + \xi_0)^2+x^2]^{K/4}}. 
\eea 
\end{subequations}
Here, $\left \langle \cos(\rho(x_1) + \rho(x_2)) \right \rangle = 0$ due to "electroneutrality" \cite{Giamarchi}.

\subsection{Spin correlation functions; easy axis} 
In the case of the EA anisotropy, the physics at energies smaller that $J_z- J_{\perp}$ is governed by $\theta \approx \pi/2$ (fast fluctuations are again neglected). At these energies the spin components become 
\be
	S^x/s =  \sin \alpha_\perp \sin \psi , \, \,
	S^y/s =- \sin \alpha_\perp \cos \psi , \, \,
	S^z/s =  \cos \alpha_\perp          . 
\ee 
Then the  transverse spin correlators are given by 
\be 
	\left \langle S^x (1) S^x(2) \right \rangle/s^2 = \left \langle S^y (1) S^y(2) \right \rangle /s^2 = \left \langle (\sin \alpha_\perp \sin \psi) (1) (\sin \alpha_\perp \sin \psi) (2) \right \rangle + {\cal O}(\theta- \pi/2, \alpha_\parallel), 
\ee 
where $(j)$ denotes $(\tau_j, x_j)$. 
Since $\psi$ and $\alpha$ are not correlated, the correlation function factorizes. The correlation function of the $\alpha_\perp$ component can be written as 
\be 
\label{sin_alpha_perp-sin_alpha_perp}
	\left \langle \sin \alpha_\perp (1) \sin \alpha_\perp (2) \right \rangle = -\tfrac 12 \left[ \left \langle \cos(2k_F (x_1 + x_2) + \alpha (1) + \alpha(2)) \right \rangle - \left \langle \cos(2k_F (x_1 - x_2) + \alpha (1) - \alpha(2)) \right \rangle \right] .
\ee 
Combining Eq. \eqref{sin_alpha_perp-sin_alpha_perp} and Eq. \eqref{basic_correlation_functions_LL} leads to 
\bea 
	\left \langle S^x (1) S^x(2) \right \rangle/s^2 &=& \tfrac 14 \cos [2 k_F (x_1 -x_2)] \left \langle \cos (\alpha(1)-\alpha(2)) \right \rangle \left \langle \cos (\psi(1) -\psi(2)) \right \rangle \nonumber \\ 
	                                   &=& \cos(2k_F x) \left ( \frac{\xi_0} {\sqrt{(\tau v_\alpha)^2+x^2}} \right )^{\frac{K_\alpha}{2}} \left ( \frac{\xi_0} {\sqrt{(\tau v_F)^2+x^2}} \right )^{\frac{1}{2}}, 
\eea 
where we introduced $x = x_1 - x_2$ and $\tau = \tau_1 - \tau_2$. 
The transverse spin correlation function of $x$ and $y$ components is  
\be 
	\label{easy_axis_transverse} 
	\left \langle S^x (1) S^y (2) \right \rangle/s^2 \sim f(\alpha_\perp) \left \langle \sin \psi (1) \cos \psi (2) \right \rangle = 0. 
\ee 
Eq. \eqref{easy_axis_transverse} shows that there is no spin rotation in $xy$-plane, see Fig. \ref{fig:spin-wave}. 
In particular, this implies that the Fourier-transform of the dynamical in-plane spin susceptibility
\be
	\label{spin_sus_ea} 
	\left \langle S^+(1) S^-(2) \right \rangle/s^2 = 2 \left \langle S^x(1) S^x(2) \right \rangle/s^2 ,%= 2 \cos(2k_F x) \left ( \frac{\xi_0} {\sqrt{(\tau v_\alpha)^2+x^2}} \right )^{\frac{K_\alpha}{2}} \left ( \frac{\xi_0} {\sqrt{(\tau v_F)^2+x^2}} \right )^{\frac{1}{2}}, 
\ee
has peaks both at $2 k_F$ and $-2k_F$.

The correlators of $S^z$ spin components are given by 
\bea 
	\left \langle S^z (1) S^z(2) \right \rangle/s^2 &=& \left \langle (\cos \alpha_\perp) (1) (\cos \alpha_\perp) (2) \right \rangle + {\cal O}(\theta- \pi/2, \alpha_\parallel) \nonumber \\ 
	                                   &=& \cos(2k_F x) \left ( \frac{\xi_0} {\sqrt{(\tau v)^2+x^2}} \right )^{\frac{K_\alpha}{2}}.
\eea 
They decay more slowly than the transverse spin correlator Eq. \eqref{easy_axis_transverse} because the $S_z$ component couples more strongly to the localized electrons. 
The correlation function between the axis and the plane $\left \langle S^z S^x \right \rangle$ vanishes. 
Thus, all cross-correlation functions are zero in the EA case.

\subsection{Spin correlation functions; easy plane} 
In the case of the EP, the asymptotics of the spin correlation functions are determined by $\theta \approx 0$, or $\theta \approx \pi$. 
Let us choose $\theta = 0$. Then the spin operators become 
\bea 
	S^x/s & = & -\cos \alpha_\perp \cos \psi + \sin \alpha_\perp \sin \psi = - \cos(\alpha_\perp + \psi) ; \\ 
	S^y/s & = & -\cos \alpha_\perp \sin \psi - \sin \alpha_\perp \cos \psi = - \sin(\alpha_\perp + \psi) ; \\ 
	S^z/s & = &  0                                                        . 
\eea 
In our notations: $\alpha_\perp = 2k_F x + \alpha$ and $\alpha \rightarrow \alpha-\psi$ in the EP case. 
Thus, the transverse spin correlation function reads as
\bea 
	\left \langle S^x (1) S^x(2) \right \rangle/s^2 &=& \left \langle [\cos(2 k_F x+\alpha)](1) [\cos(2 k_F x+\alpha)](2) \right \rangle \nonumber \\ 
	                                   &=& \cos(2k_F x) \left ( \frac{\xi_0} {\sqrt{(\tau v_\alpha')^2+x^2}} \right )^{K_{\alpha}'/2} .
\eea 
Due to $SO(2)$-symmetry in the $x$-$y$-plane, this is the same as the $\left \langle S^y S^y \right \rangle$ correlation function. 
The transverse spin rotation correlation function is  
\be 
\label{easy_plane_transverse} 
	\left \langle S^x (1) S^y (2) \right \rangle/s^2 = \sin (2 k_F x) \left ( \frac{\xi_0}{\sqrt{(\tau v_\alpha')^2 + x^2}} \right )^{K_{\alpha}'/2} .
\ee 
Eq. \eqref{easy_plane_transverse} reveals the spin rotation (helical configuration) in the EP case, see Fig. \ref{fig:spin-wave}. 
Contrary to the EA case, the Fourier transform of the dynamical in-plane spin susceptibility
\be
\label{spin_sus_ep}
	\left \langle S^+(1) S^-(2) \right \rangle/s^2 = 2 \left ( \left \langle S^x(1) S^x(2) \right \rangle - \ri \left \langle S^x S^y \right \rangle \right )/s^2 = \exp (- \ri 2 k_F x) \left ( \frac{\xi_0}{\sqrt{(\tau v_\alpha')^2 + x^2}} \right )^{K_{\alpha}'/2} 
\ee
has a peak only at $2 k_F$.
The longitudinal spin correlator $\left \langle S^z S^z \right \rangle$ is zero in our accuracy (at fixed $\theta=0$, $\alpha_\parallel=0$).

\subsection{Order parameter} 
We have shown that the low energy spin excitations of the EA case are planar spin oscillations, whereas in the EP case the spins form a helix, see Fig. \ref{fig:spin-wave}. 

The transverse spin correlation function $\left \langle S^x (1) S^y (2) \right \rangle$, which reflects rotations of the spins, is zero in the non-helical phase (EA), but nonvanishing in the helical one (EP). Thus we suggest to use it as an order parameter. In analogy with antiferromagnetic ordering \cite{Fradkin}, we define the two-point order parameter 
\be 
	{\cal A}_c = \epsilon_{abc} \left \langle S^a (1) S^b (1+\xi_0) \right \rangle, 
\ee 
which is non-vanishing only in the helical phase, where there is a low-energy helical mode propagating within the dense chain of the magnetic impurities. 
 
%This also explains $\left \langle S^z S^z \right \rangle = 0$ in the easy plane case. 

\section{Conclusion} 
\label{sec:Conclusion}
Low-energy properties of an anisotropic Kondo chain away from half-filling are governed either by the Kondo physics or by the backscattering of the fermions. The latter dominates if the concentration of spins is sufficiently large and if there are sufficiently strong repulsive electron-electron interactions. The dominance of backscattering leads to the opening of a gap in the fermionic spectrum, Eq. \eqref{GapsPM}, which suppresses the emergence of the Kondo physics. Depending on the anisotropy of the exchange interaction, the backscattering processes  may either  lead to a formation of  charge and spin density waves (EA anisotropy), Eq. \eqref{LagrEA}, or to helical low energy modes (EP anisotropy), Eq. \eqref{LagrEP2}. The latter ones  appear if the helical ${\mathbb Z}_2$-symmetry is spontaneously broken. We have shown that the order parameter characterizing the corresponding quantum phase transition is the average of the vector product of neighboring spins ${\cal A}_c = \epsilon_{abc} \left \langle S^a (1) S^b (1+\xi_0) \right \rangle$. The helical nature of the modes is also manifest in the assymetry between $+2k_F$ and $-2k_F$ peaks in the  in-plane spin susceptibility $\left \langle S^+ S^- \right \rangle$, Eq. \eqref{spin_sus_ep}. The ideal charge transport supported by the gapless helical modes is robust: it remains ballistic even if a weak random potential of static impurities is present. This protection requires the spin U(1) symmetry and it exists up to a parametrically large scale, see Eq. \eqref{eq:Localization_radii}. We have shown that short-range electron-electron interactions mix the two helical sectors, but cannot gap out any low-energy modes, such that for weak interactions the qualitative description by helical modes remains valid.

Even though the helical modes may be reminiscent of the edge modes of topological insulators, we emphasize that, in our case, they emerge from many body interaction effects in one dimension. Experimentally, the helical modes could be detected in samples exhibiting one-dimensional structure with spin impurities. Promising candidates are ladder-type $Fe$-selenides, where almost completely filled bands of electrons might serve as spin impurities \cite{Dagotto_PRL}, or single-wall carbon nanotubes \cite{PhysRevLett.95.017401}. Since the advent of the cleaved edge overgrowth method \cite{Pfeiffer1997817}, quantum wires on the edge of GaAs heterostructures are also viable candidates. 

Usually, one cannot control the anisotropy of real materials. Therefore, one needs an
experimental evidence that the charge transport in a given system with the dense array
of the Kondo impurities is supported by modes with the broken helical symmetry. The 
cleanest signature could be provided by the local spin susceptibility (Eqs. \eqref{spin_sus_ea} 
and \eqref{spin_sus_ep}), which clearly provides a smoking gun signature for helical order. 
It seems possible that the local spin susceptibility is experimentally accessible through nitrogen-vacancy based STM 
measurements if the Kondo array is made as a one-dimensional wire \cite{Loss_spin_sus}. Another
simplest experimental signature of the helical phase might be frequency-resolved charge transport. 
We remind the readers that charge is carried by the collective mode $\alpha$ (EA), or the collective 
mode $\alpha$ and the helical fermion (EP) with the velocity of $\alpha$ being always small 
(Eqs. \eqref{Compr-a}  and \eqref{velocitiesEP}). If a finite and sufficiently clean sample is connected 
to leads adiabatically, its dc conductance remains ideal, $2e^2/h$ 
\cite{MaslovStone}. However, the frequency resolved conductance is expected to show
a substantial decrease at $\omega_c \sim 1 / t_{\rm Th}^{(\alpha)}$; where $t_{\rm Th}^{(\alpha)}
\sim L / v_\alpha$ is the Thouless time associated with the mode $\alpha$. Since $\alpha$-modes 
are very slow $ \omega^{-1}_c $ is small. For frequencies larger than $ \omega^{-1}_c $, the slow 
collective modes cannot contribute and the conductance drops either to zero (EA) or to $e^2/h$ (EP). 
The latter jump would confirm that the system is in the helical phase which is robust against localization 
effects. The theory of the frequency dependent conductance of the Kondo chain requires further 
theoretical work. 

%Detection of the helical mode could be achieved by measuring the in-plane spin susceptibility $\left \langle S^+ S^- \right \rangle$, or the parametric suppression of the localization radius.

\begin{acknowledgments}
A.M.T. acknowledges the hospitality of Ludwig Maximilians University where part of this work was done.
A.M.T. was supported by the U.S. Department of Energy (DOE), Division of Materials Science, under
Contract No. DE-AC02-98CH10886. 
O.M.Ye. acknowledges support from the DFG through SFB TR-12, and the Cluster of Excellence, Nanosystems Initiative Munich. 
D.H.S. is supported through by the DFG through
the  Excellence  Cluster  “Nanosystems  Initiative  Munich”, SFB/TR 12 and SFB 631.
We are grateful to Vladimir Yudson and Igor Yurkevich for useful discussions.
\end{acknowledgments}

\begin{appendix} 
\section{Derivation of the low-energy Lagrangian} 
In this section we give a short derivation of the form of the electron-spin interactions in terms of the fast and slow angular variables ($\alpha_\parallel$, $\alpha_\perp$, $\theta$, and $\psi$).
Thus, consider the interaction term  
\be 
   {H}_{int} = \sum_m J_a \, \hat{c}^\dagger_m \, \hs^a \hat{S}^a(m) \, \hat{c}_m. 
\ee 
Using the representation of the fermions in terms of left- and rightmovers, Eq. \eqref{fermions}, this term splits into forward and backward scattering contributions 
\be 
   {H}_{int} =  H_{\rm forward} +  H_{\rm backward},
\ee 
\be 
   {H}_{\rm forward} = \sum_m J^f_a \, \hat{R}^\dagger_m \, \hs^a \hat{S}^a(m) \, \hat{R}_m 
                  +\sum_m J^f_a \, \hat{L}^\dagger_m \, \hs^a \hat{S}^a(m) \, \hat{L}_m 
\ee 
\be 
   {H}_{\rm backward} = 
                  +\sum_m J^b_a \re^{ 2 \ri k_F x} \, \hat{R}^\dagger_m \, \hs^a \hat{S}^a(m) \, \hat{L}_m 
                  +\sum_m J^b_a \re^{-2 \ri k_F x} \, \hat{L}^\dagger_m \, \hs^a \hat{S}^a(m) \, \hat{R}_m, 
\ee 
where the superscript $f$ ($b$) denotes forward (backward) scattering contributions. 
Using the low-energy spin $S_{LE} \parallel \vec e_3'$ and taking the dense impurity limit, we obtain
\bea 
\label{L_bs_start}
	{\cal L}^{(bs)}_{\rm int} &= s \rho_s \re ^{2 \ri k_F x} R^\dagger \big \{ \frac{J^b_\perp}{2}  \left [\right . &+ \re^{i\psi} (-\cos \alpha_\parallel \cos \alpha_\perp \cos \theta - \ri \cos \alpha_\parallel \sin \alpha_\perp + \sin \alpha_\parallel \sin \theta) \hat \sigma^- \nonumber \\ 
	               &                                 &\left .+ \re^{-i\psi} (-\cos \alpha_\parallel \cos \alpha_\perp \cos \theta + \ri \cos \alpha_\parallel \sin \alpha_\perp + \sin \alpha_\parallel \sin \theta) \hat \sigma^+  \right ]  \nonumber \\ 
						&                                 &+ J^b_z \hat \sigma_z (\sin \alpha_\parallel \cos \theta + \cos \alpha_\parallel \cos \alpha_\perp \sin \theta) \big \} L + H.c. .
\eea 
This expresses the back-scattering part of the electron-spin interaction in terms of the angular variables and the fermions. To obtain the low-energy part, we first 
shift $\alpha_\perp \rightarrow \alpha(x) + 2 k_F x$. Then,
neglecting all quickly oscillating terms ($\sim \re^{4 \ri k_F x}$), Eq. \eqref{L_bs_start} reduces to
\be 
   {\cal L}_{int}^{\rm (sl) (bs)} = \frac{s \cos(\alpha_\parallel) \rho_s}{2} 
             R^\dagger \Bigl\{ 
                 J_\perp \left[ \re^{\ri \psi} \sin^2 \! \left( \frac{\theta}{2} \right)\hs^-  \!\! - 
                                       \re^{-\ri \psi} \cos^2 \! \left( \frac{\theta}{2} \right)\hs^+ 
                               \right] 
                                  + J_z \sin(\theta)\hs^z \Bigr\} L \re^{-\ri \alpha} + H.c. ; \quad \tilde{s} \equiv s \cos(\alpha_\parallel) 
\ee 
The forward-scattering part of the action is obtained by following the same procedure with $H_{\rm forward}$:
\be 
	{\cal L}_{int}^{\rm (sl) (fs)} = \frac{s \sin(\alpha_\parallel) \rho_s}{2} R^\dagger \left \{ J_\perp^f \sin \theta [\re^{\ri \psi} \sigma^- + \re^{-\ri \psi} \sigma^+] + 2 J_z^f \cos \theta \sigma^z \right \} R + (R \rightarrow L) 
\ee

\section{Bosonization and the RG equations} 
\label{sec:App_RG} 
Here we briefly remind readers of the bosonization identity used throughout, and the derivation of the RG equations. We only derive one RG equation explicitely, but the other RG equations may be obtained by the same procedure.

The bosonization formula is 
\be 
\label{eq:bosonization}
	\Psi_{r \sigma} = \frac{1}{\sqrt{2\pi \alpha}} U_{\sigma} \re^{-\ri r k_F x} \re^{-\frac{1}{\sqrt 2}\ri [r \Phi_c-\Theta_c+\sigma(r \Phi_s-\Theta_s)]}, 
\ee 
where $\Phi_c$ ($\Phi_s$) and $\Theta_c$ ($\Theta_s$) are dual fields belonging to the charge (spin) density wave, $r$ distinguishes right- and left-moving and $\sigma$ is the spin. The Klein factors $U_\sigma$ are real coordinate independent fermionic operators obeying the anticommutation relations $\{U_{\s},U_{\s'}\} = \delta_{\s,\s'}$.
 
After bosonization Eq. \eqref{eq:bosonization}, the electron-spin interaction contains the terms 
\begin{flalign} 
	{\cal L}^f_z     &:= J^f_z     S_z ( R^\dagger \sigma_z R + L^\dagger \sigma_z L )      &=& - \sqrt 2 S_z \frac{J^f_z}{\pi} \partial_x \Phi_s & \nonumber \\ 
	{\cal L}^f_-     &:= J^f_\perp S_- ( R^\dagger \sigma_+ R + L^\dagger \sigma_+ L )      &=&     S_- \frac{J^f_\perp}{\pi \xi_0} \exp (-\sqrt 2 \ri \Theta_s)  \left (\exp(\sqrt 2 \ri \Phi_s) + \exp(-\sqrt 2 \ri \Phi_s) \right ) & \nonumber \\ 
	{\cal L}^f_+     &:= J^f_\perp S_+ ( R^\dagger \sigma_- R + L^\dagger \sigma_- L )      &=&     S_+ \frac{J^f_\perp}{\pi \xi_0} \exp ( \sqrt 2 \ri \Theta_s)  \left (\exp(-\sqrt 2 \ri \Phi_s) + \exp(\sqrt 2 \ri \Phi_s) \right ) & \nonumber \\ 
	{\cal L}^b_z     &:= J^b_z     S_z ( R^\dagger \sigma_z L + L^\dagger \sigma_z R )      &=&     S_z \frac{J^f_z    }{2 \pi \xi_0} \exp (-2 \ri k_F x) \exp (\sqrt 2 \ri \Phi_c)  \left (\exp(\sqrt 2 \ri \Phi_s) - \exp(-\sqrt 2 \ri \Phi_s) \right ) + H.c. & \nonumber \\ 
	{\cal L}^b_-     &:= J^b_\perp ( S_-  R^\dagger \sigma_+ L + S_+ L^\dagger \sigma_- R ) &=&     S_- \frac{J^f_\perp}{\pi \xi_0} \exp (-2 \ri k_F x) \exp  \left (\sqrt 2 \ri (\Phi_c - \Theta_s)  \right ) + H.c. & \nonumber \\ 
\label{eq:bosonized_vertices} 
	{\cal L}^b_+     &:= J^b_\perp ( S_+  R^\dagger \sigma_- L + S_- L^\dagger \sigma_+ R ) &=&     S_+ \frac{J^f_\perp}{\pi \xi_0} \exp (-2 \ri k_F x) \exp  \left (\sqrt 2 \ri (\Phi_c + \Theta_s)  \right ) + H.c. & 
\end{flalign} 
The flow of the coupling constants is obtained by integrating out high energy modes. To do so, one must split $\Phi_\alpha$, $\Theta_\alpha$ and $S_\beta$ into fast (superscript $>$) and slow (superscript $<$) modes: 
\be 
	\Phi_\alpha = \Phi_\alpha^< + \Phi_\alpha^>, \quad \Theta_\alpha = \Theta_\alpha^< + \Theta_\alpha^>, \quad S_\beta = S_\beta^< + S_\beta^>. 
\ee 
The measure of the path integral splits into fast and slow modes as well. We then perform the integral over the fast modes in a perturbative series in $J$ and reexponentiate the result. The first order in $J$ leads to the one-loop RG equations.  
As in the bosonization treatment of the Kondo impurity, we will treat the spins as constant during the RG flow.
Thus, we need to compute 
\be 
	\int {\cal D} \{\Phi, \Theta\} \exp \left (-S_{LL}[\Phi, \Theta] - \int d\tau dx J_a S_a f_a(\Phi, \Theta) \right ), 
\ee 
where $S_{LL}$ is the Luttinger liquid action for $\Phi$ and $\Theta$ and $f_a$ is a function which can be read off from \eqref{eq:bosonized_vertices}. Note that there is space-time UV cutoff $\xi_0$ (or equivalently an energy-momentum cutoff $\Lambda$). Let us consider as an example the term proportional to $J^b_z$:
\bea 
	&   &\int {\cal D} \{\Phi^>, \Theta^>\} \exp \left (-S_{LL}[\Phi^>, \Theta^>]\right) \int d\tau dx J^b_z S^<_z f^b_z \left(\Phi^> + \Phi^<, \Theta^> + \Theta^< \right) \nonumber \\ 
	& = &\int d\tau dx J^b_z S^<_z \int {\cal D} \{\Phi^>, \Theta^>\} \exp \left(-S_{LL}[\Phi^>, \Theta^>] \right) \frac{1}{2 \pi \xi_0} \exp (-2 \ri k_F x) \exp \left(\sqrt 2 \ri (\Phi^>_c+\Phi^<_c) \right) \nonumber \\ 
	& & \quad \quad \quad \times \left (\exp \left (\sqrt 2 \ri (\Phi^>_s+\Phi^<_s)\right ) - \exp \left(-\sqrt 2 \ri (\Phi^>_s+\Phi^<_s)\right )\right ) + H.c. 
\eea 
The components $\Phi^>$ ($\Theta^>$) and $\Phi^<$ ($\Theta^<$) are of high and low energy, such that the energy of $\Phi^>$ ($\Theta^>$) lies in the interval $[\Lambda',\Lambda]$. 
Using the equalities $\left \langle \re^{\sqrt 2 \ri \Phi^>_s} \right \rangle_> = (\Lambda'/\Lambda)^{K_s/2}$ and $\left \langle \re^{\sqrt 2 \ri \Theta^>_s} \right \rangle_> = (\Lambda'/\Lambda)^{1/(2K_s)}$, we can perform the average over fast modes. This yields 
\bea 
	&   &\int {\cal D} \{\Phi^>, \Theta^>\} \exp \left (-S_{LL}[\Phi^>, \Theta^>] \right) \int d\tau dx J^b_z S^<_z f^b_z \left(\Phi^> + \Phi^<, \Theta^> + \Theta^< \right ) \nonumber \\ 
	\label{eq:RGbz} 
	& = &\int d\tau dx J^b_z S^<_z \left( \frac{\Lambda'}{\Lambda} \right)^{\tfrac 12 (K_s + K_c)} f^b_z(\Phi^<, \Theta^<) 
\eea 
Since the cutoff was changed from $\Lambda'$ to $\Lambda$, we need to rescale $x$ and $\tau$ to recover the original expression. Reexponentiating \eqref{eq:RGbz} yields 
\be 
\label{scaling_Appendix_example}
	J^b_z (\Lambda') = J^b_z (\Lambda) \left( \frac{\Lambda'}{\Lambda} \right)^{\tfrac 12 (K_s + K_c) -2} 
\ee 
The RG equation is obtained expressing Eq. \eqref{scaling_Appendix_example} as a differential equation in the parametrization $\Lambda' = \Lambda \re^{-l-dl}$, where $dl$ is an infinitesimal number:
\be 
	\partial_l J_z^b = \left [ \tfrac 12 (K_s + K_c) - 2 \right ] J_z^b.
\ee

\section{The shift of the angles in the easy axis case}
\label{sec:App_Shift} 
We present a short, alternative derivation of the action after the shift eliminiating the angles $\alpha$ and $\psi$ from the interaction vertices, Eq. \eqref{Ltotal}. This proof is based on abelian bosonization. Upon bosonization, Eq. \eqref{eq:bosonization}, the free part of the Lagrangian are a spin and charge Tomonaga-Luttinger liquid:
\be
\label{TL_Appendix}
	{\cal L} = {\cal L}_{TL, \rm dual}[\Phi_c, \Theta_c] + {\cal L}_{TL, \rm dual}[\Phi_s, \Theta_s],
\ee
with
\be
	{\cal L}_{TL, \rm dual} [\Phi_a, \Theta_a] = - \frac{\ri}{\pi} \partial_x \Theta_a \partial_\tau \Phi_a + \frac{1}{2\pi} \left ( u K (\partial_x \Theta_a)^2 + \frac{u}{K} (\partial_x \Phi_a)^2 \right ).
\ee
We use a description in terms of fields $\Phi$ and their duals $\Theta$. 
The shift Eq. \eqref{fermion-trafo} is in bosonic language
\be
	\Phi_c \rightarrow \Phi_c + \alpha/\sqrt 2, \, \Theta_s \rightarrow \Theta_s - \psi/\sqrt 2.
\ee
Performing this shift also in the Tomonaga-Luttinger liquid Eq. \eqref{TL_Appendix}, we obtain the new terms of the form
\be
\label{L_TL_mixing}
	{\cal L}_{\rm mixing} \sim -\ri \partial_\tau \alpha \partial_x \Theta_c - \partial_x \alpha \partial_x \Phi_c - \ri \partial_\tau \psi \partial_x \Phi_x - \partial_x \psi \partial_x \Theta_s.
\ee
and terms of the type
\be
\label{L_alpha_psi_TL}
	{\cal L}_{TL, \rm dual} [\alpha/\sqrt 2, \Theta_c] + {\cal L}_{TL, \rm dual} [\Phi_s, \psi/\sqrt 2].
\ee
Since after bosonization spatial derivatives of $\Phi_{c/s}$ ($\Theta_{c/s}$) correspond to the charge/spin density (current), Eq. \eqref{L_TL_mixing} contains precisely the terms of Eq. \eqref{L_mixing}, and may be neglected by the same arguments.
After averaging over the dual fields $\Theta_c$ and $\Phi_s$, Eq. \eqref{L_alpha_psi_TL} is the same as the Tomonaga-Luttinger anomaly Eq. \eqref{action_TL}. 
We thus have obtained the same expression as in the main text, without explicitely using the Tomonaga-Luttinger anomaly.

\section{Accounting for interactions} 
In this section we show how to obtain Eq. \eqref{L_int_ep}. 
We start from the bosonized Lagrangian of interacting electrons
\be 
\label{Appendix_LL} 
	{\cal L} =- \frac{\ri}{\pi} \partial_x \Theta_c \partial_\tau \Phi_c + \frac{1}{2 \pi} \left (u_c K_c \left (\partial_x \Theta_c \right )^2 + \frac {u_c}{K_c} \left (\partial_x \Phi_c \right )^2 \right ) 
	          - \frac{\ri}{\pi} \partial_x \Theta_s \partial_\tau \Phi_s + \frac{1}{2 \pi} \left (u_s K_s \left (\partial_x \Theta_s \right )^2 + \frac {u_s}{K_s} \left (\partial_x \Phi_s \right )^2 \right ) 
\ee 
In order to rewrite Eq. \eqref{Appendix_LL} in terms of helical fields, we define
\begin{subequations}
\bea 
\label{definition_helical} 
	\Phi_{h_1} &= \frac{1}{\sqrt{2}} (\Phi_c - \Theta_s), \quad \Theta_{h_1} = \frac{1}{\sqrt{2}} (\Theta_c - \Phi_s) \\ 
	\Phi_{h_2} &= \frac{1}{\sqrt{2}} (\Phi_c + \Theta_s), \quad \Theta_{h_2} = \frac{1}{\sqrt{2}} (\Theta_c + \Phi_s) .
\eea 
\end{subequations}
This choice stems from the identities
\bea
	\rho_\downarrow^R = \frac{\sqrt{2}}{\pi} \partial_x ( \Theta_c -\Phi_c-(\Theta_s-\Phi_s)) \nonumber \\ 
	\rho_\uparrow^L   = \frac{\sqrt{2}}{\pi} \partial_x (-\Theta_c -\Phi_c- \Theta_s-\Phi_s ) .
\eea
If there are no particles of one specific helical sector (e.g. $R_\downarrow$ and $L_\uparrow$), then both of these densities should vanish. This is guaranteed if there are no fluctuations in $\Phi_{h2}$ and $\Theta_{h2}$. Thus, the fields $\Phi_{h2}$ and $\Theta_{h2}$ correspond to the helical sector containing $R_\downarrow$ and $L_\uparrow$.

Inserting Eq. \eqref{definition_helical} into Eq. \eqref{Appendix_LL}, we obtain 
\bea 
	2{\cal L} =&-  \frac{\ri}{\pi} \partial_x (\Theta_{h_1}+\Theta_{h_2}) \partial_\tau (\Phi_{h_1}+\Phi_{h_2}) + \frac{1}{2 \pi}  \left (u_c K_c  \left (\partial_x (\Theta_{h_1}+\Theta_{h_2}) \right )^2 + \frac {u_c}{K_c}  \left (\partial_x (\Phi_{h_1}+\Phi_{h_2}) \right )^2 \right ) \nonumber \\ 
	          & -  \frac{\ri}{\pi} \partial_x (-\Phi_{h_1}+\Phi_{h_2}) \partial_\tau (-\Theta_{h_1}+\Theta_{h_2}) + \frac{1}{2 \pi}  \left (u_s K_s (\partial_x  \left (-\Phi_{h_1}+\Phi_{h_2} \right ))^2 + \frac {u_s}{K_s}  \left (\partial_x (-\Theta_{h_1}+\Theta_{h_2}) \right )^2 \right )\\ 
	          =&- 2\frac{\ri}{\pi} \partial_x (\Theta_{h_1}) \partial_\tau (\Phi_{h_1}) + \frac{1}{2 \pi}  \left ( \left (u_c K_c+\frac {u_s}{K_s} \right ) (\partial_x \Theta_{h_1})^2 +  \left (\frac {u_c}{K_c}+u_s K_s \right ) (\partial_x \Phi_{h_1})^2 \right ) \nonumber \\ 
	          & - 2\frac{\ri}{\pi} \partial_x (\Theta_{h_2}) \partial_\tau (\Phi_{h_2}) + \frac{1}{2 \pi}  \left ( \left (u_c K_c+\frac{u_s}{K_s} \right ) (\partial_x \Theta_{h_2})^2 +  \left (\frac {u_c}{K_c}+u_s K_s \right ) (\partial_x \Phi_{h_2})^2 \right ) \nonumber \\ 
	          &+ \frac{1}{2 \pi}  \left (2 \left (u_c K_c-\frac{u_s}{K_s} \right ) \partial_x \Theta_{h_2} \partial_x \Theta_{h_1} + 2 \left (\frac {u_c}{K_c}-u_s K_s \right ) \partial_x \Phi_{h_2} \partial_x \Phi_{h_1} \right ) 
\eea 
The shift Eq. \eqref{trafo_ep}, which keeps the second helical sector invariant, corresponds to $\Phi_{h_1} \rightarrow \Phi_{h_1} + \alpha/2$. After neglecting couplings between gapless modes and derivatives of the first helical sector, we find in addition to the free part ${\cal L}_{\rm TL}$ of $\alpha$
\bea 
\label{Appendix_L_Helical}
	{\cal L}=& - \frac{\ri}{\pi} \partial_x \Theta_{h_1} \partial_\tau \Phi_{h_1} + \frac{1}{4 \pi}  \left ( \left (u_c K_c+\frac {u_s}{K_s} \right ) (\partial_x \Theta_{h_1})^2 +  \left (\frac {u_c}{K_c}+u_s K_s \right ) (\partial_x \Phi_{h_1})^2 \right ) \nonumber \\ 
	          &- \frac{\ri}{\pi} \partial_x \Theta_{h_2} \partial_\tau \Phi_{h_2} + \frac{1}{4 \pi}  \left ( \left (u_c K_c+\frac{u_s}{K_s} \right ) (\partial_x \Theta_{h_2})^2 +  \left (\frac {u_c}{K_c}+u_s K_s \right ) (\partial_x \Phi_{h_2})^2 \right ) \nonumber \\ 
	          &+ \frac{1}{2\pi} \left (\frac {u_c}{K_c}-u_s K_s \right ) \partial_x \Phi_{h_2} \partial_x \alpha. 
\eea 
Introducing  
\be 
	\tilde K = \sqrt{\frac{u_c K_c + \frac {u_s}{K_s}}{\frac{u_c}{K_c} + u_s K_s}}, \quad \tilde u = \frac{1}{4} \sqrt{u_c^2 + u_s^2 + u_c u_s K_c K_s + \frac{u_c u_s}{K_c K_s}}, 
\ee 
Eq. \eqref{Appendix_L_Helical} may be written as 
\bea 
	{\cal L}=& - \frac{\ri}{\pi} \partial_x \Theta_{h_1} \partial_\tau \Phi_{h_1} + \frac{1}{2 \pi}  \left (\tilde u \tilde K (\partial_x \Theta_{h_1})^2 + \tilde u \frac{1}{\tilde K} (\partial_x \Phi_{h_1})^2 \right ) \nonumber \\ 
	          &- \frac{\ri}{\pi} \partial_x \Theta_{h_2} \partial_\tau \Phi_{h_2} + \frac{1}{2 \pi}  \left (\tilde u \tilde K (\partial_x \Theta_{h_2})^2 + \tilde u \frac{1}{\tilde K} (\partial_x \Phi_{h_2})^2 \right ) \nonumber \\ 
	          &+ \frac{1}{2\pi}(\frac {u_c}{K_c}-u_s K_s) \partial_x \Phi_{h_2} \partial_x \alpha. 
\eea

\section{Non-Gaussianities in the effective disorder}
\label{Appendix_effective_disorder}
In this Appendix, we demonstrate that the higher moments of the effective disorder $g_{\rm eff}$ distribution function in the alternative approach to disorder are of higher order in $\frac{{\cal D}}{v_F m} \ll 1$. Thus, in our accuracy, we may safely neglect the non-Gaussianities of the effective disorder.

We have assumed that the distribution of the $2k_F$ Fourier components of the original disorder potential is Gaussian, however the distribution of $g_{\rm eff}(x)$ is not Gaussian.  
To investigate the effect of the non-Gaussianity of the distribution function of the effective disorder $g_{\rm eff}$, we consider its moments. The first moment is zero:
\be
	\left \langle g_{\rm eff}(x) \right \rangle_{\rm dis} \sim \left \langle \frac{1}{v_F} \int dy g(x+y/2) g(x-y/2) \re^{-m \vert y \vert/v_F} \right \rangle_{\rm dis} = 0,
\ee
because $g$ is distributed according to the GUE. The second moment is given by
\be
	\left \langle g_{\rm eff}(x) g_{\rm eff}(x') \right \rangle_{\rm dis} \sim \left \langle \frac{1}{v_F^2} \int dy d\tilde y g(x+y/2) g(x-y/2) g(x'+\tilde y/2) g(x'-\tilde y/2) \re^{-m (\vert y \vert+\vert \tilde y \vert )/v_F} \right \rangle_{\rm dis} = 0,
\ee
and
\bea
	\left \langle g_{\rm eff}(x) g^*_{\rm eff}(x') \right \rangle_{\rm dis} \sim& \frac{1}{v_F^2} \Big \langle \displaystyle{\int } dy d\tilde y & g(x+y/2) g(x-y/2) g^*(x'+\tilde y/2) g^*(x'-\tilde y/2) \re^{-m (\vert y \vert+\vert \tilde y \vert )/v_F} \Big \rangle_{\rm dis} \nonumber    \\
	                                                           \sim& \frac{{\cal D}^2}{v_F^2} \displaystyle{\int } dy d\tilde y &(\delta(x+y/2-x'+\tilde y/2) \delta(x-y/2-x'-\tilde y/2)\nonumber \\
																				  & & +\delta(x+y/2-x'-\tilde y/2) \delta(x-y/2-x'+\tilde y/2)) \re^{-m (\vert y \vert+\vert \tilde y \vert )/v_F} \nonumber  \\
																				  \sim & \frac{{\cal D}^2}{v_F m} \delta(x-x').
\eea

Higher moments contain additional contractions, reflecting the non-Gaussianity of the distribution of $g_{\rm eff}$. As an example, consider the fourth moment
\bea
	\left \langle g_{\rm eff}(x) g_{\rm eff}(y) g^*_{\rm eff}(z) g^*_{\rm eff}(w) \right \rangle_{\rm dis} \sim& \frac{1}{v_F^4} 
	                                                            \Big \langle \displaystyle{\int } dx' dy' dz' dw' & g(x+x'/2) g(x-x'/2) g(y+y'/2) g(y-y'/2)  \nonumber \\
	                                                          &                              & g^*(z+z'/2) g^*(z-z'/2) g^*(w+w'/2) g^*(w-w'/2) \nonumber \\ 
																				&										&	 \re^{-m (\vert x' \vert+\vert y' \vert+\vert z' \vert+\vert w' \vert)/v_F} \Big \rangle_{\rm dis} . 
\eea
There are two distinct kinds of contractions: Gaussian ones (contracting e.g. $\langle g(x+x'/2) g^*(z+z'/2) \rangle$, $\langle g(x-x'/2) g^*(z-z'/2) \rangle$, $\langle g(y+y'/2) g^*(w+w'/2) \rangle$, and $\langle g(y-y'/2) g^*(w-w'/2) \rangle$) and non-Gaussian ones, e.g. contracting $\langle g(x-x'/2) g^*(z-z'/2) \rangle$, $\langle g(x+x'/2) g^*(w+w'/2) \rangle$, $\langle g(y+y'/2) g^*(z+z'/2) \rangle$, and $\langle g(y-y'/2) g^*(w-w'/2) \rangle$. The latter yields:
\bea
\label{app_disorder_nonGauss}
	\left \langle g_{\rm eff}(x) g_{\rm eff}(y) g^*_{\rm eff}(z) g^*_{\rm eff}(w) \right \rangle_{\rm dis} \supset& \frac{{\cal D}^4}{v_F^4} 
	                                                                    \displaystyle{\int } dx' dy' dz' dw' & \delta(x-x'/2-z+z'/2) \delta(x+x'/2-w-w'/2) \nonumber \\
	                                                          &                              & \delta(y+y'/2-z-z'/2) \delta(y-y'/2-w+w'/2) \nonumber \\ 
																				&										&	 \re^{-m (\vert x' \vert+\vert y' \vert+\vert z' \vert+\vert w' \vert)/v_F} \nonumber \\
	                                                                                           \sim   & \frac{{\cal D}^4}{v_F^4} 
	                                                                    \displaystyle{\int } dx' dy' dz' dw' & \delta(z' - y+x-y'/2-x'/2) \delta(w' - x+y-y'/2-x'/2) \nonumber \\
	                                                          &                              & \delta(x' - 2w +2z -y') \delta(z+w-x-y) \nonumber \\ 
																				&										&	 \re^{-m (\vert x' \vert+\vert y' \vert+\vert z' \vert+\vert w' \vert)/v_F} \nonumber \\
	                                                                                           \sim   & \frac{{\cal D}^4}{v_F^4} 
	                                                                    \delta(z+w-x-y) \displaystyle{\int } dy' & \re^{-m (\vert 2w-2z+y' \vert+\vert y' \vert+\vert y'-2z+2y \vert+\vert 2w-2y+y' \vert)/v_F}. 
\eea
In addition to the phase space factor of $v_F/m$, we obtain an exponential suppression of lengths $(w-z)$ etc. larger than $v_F/m$. The leading order for large distances may be extracted by
formally taking the limit $m \rightarrow \infty$. The exponential may then be approximated by a $\delta$-function: $\delta(x) = \lim_{m\rightarrow \infty} (m/v_F) \exp(-m \vert x \vert/v_F)$. Note that in the case of multiple terms in the exponent some of them might be spurious, i.e. $\exp(-m (\vert x \vert + \vert x \vert)/v_F) \sim (v_F/m) \delta(x)$. Taking this into account the large-distance limit of Eq. \eqref{app_disorder_nonGauss} leads to
\be
\label{app_disorder_nonGauss_correct}
	\left \langle g_{\rm eff}(x) g_{\rm eff}(y) g^*_{\rm eff}(z) g^*_{\rm eff}(w) \right \rangle_{\rm dis} \sim  \frac{{\cal D}^4}{v_F^{4}} \delta(z+w-x-y) \frac{v_F^3}{m^3} \delta(z-w) \delta(y-w)
\ee

%Naively, Eq. \eqref{app_disorder_nonGauss} thus leads to
%\be
%\label{app_disorder_nonGauss_naive}
%	\left \langle g_{\rm eff}(x) g_{\rm eff}(y) g^*_{\rm eff}(z) g^*_{\rm eff}(w) \right \rangle_{\rm dis} 
%	                                                                                           \sim   \frac{{\cal D}^4}{v_F^4} 
%	                                                                    \delta(z+w-x-y) \frac{v_F^4}{m^4} \delta(w-z) \delta(x-y) \delta(w-y),
%\ee
%which is ill-defined (recall that the description by a $\delta$-function is only valid in the limit $m\rightarrow \infty$). The reason is that we have approximated $\exp(-m (\vert x \vert + \vert x \vert)/v_F)$ by $(v_F/m)^2 \delta(x) \delta(x)$, which is wrong. It should be $\exp(-m (\vert x \vert + \vert x \vert)/v_F) \sim v_F/m \delta(x)$. Correctly perfoming the large-distance limit leads to
%\be
%\label{app_disorder_nonGauss_correct}
%	\left \langle g_{\rm eff}(x) g_{\rm eff}(y) g^*_{\rm eff}(z) g^*_{\rm eff}(w) \right \rangle_{\rm dis} \sim  \frac{{\cal D}^4}{(v_F)^{4}} \delta(z+w-x-y) \frac{v_F^3}{m^3} \delta(z-w) \delta(y-w)
%\ee

Higher moments are suppressed in a similar fashion. Thus, we have proven that the non-Gaussian contributions are supressed by at least the factor $\frac{{\cal D}^2}{(v_F m)^2}$.
%Analogously, the $2n$-th moment contain two contributions: a Gaussian part of the form $\frac{{\cal D}^{2n}}{(v_F m)^{n}} \delta^{n}(\dots)$, where the dots stand for all contractions of the positions, and non-Gaussian parts 
%which are of the form $\frac{{\cal D}^{2n}}{v_F^{2n}} \frac{v_F^{n-r}}{m^{n-r}} \delta^{n-r}(\dots)$, with $r \geq 1$.

\end{appendix} 
\bibliography{Bibliography} 
 
\end{document}